\documentclass[superscriptaddress,amsmath,amssymb,aps,twocolumn,praplied,notitlepage]{revtex4-2}
\usepackage{amsmath,amsfonts,amssymb,amsthm,graphics,graphicx,epsfig,bbm}
\usepackage[colorlinks=true,citecolor=blue,linkcolor=blue,urlcolor=blue]{hyperref}
\usepackage[usenames]{color}
\usepackage{graphicx}
\usepackage{subfigure}
\usepackage{amsmath}
\usepackage{epsfig}
\usepackage{dcolumn}
\usepackage{bm}
\usepackage{color}
\usepackage{times}
\usepackage{epstopdf}
\usepackage{amssymb}
\usepackage{amstext}
\usepackage{latexsym}
\usepackage{hyperref}
\usepackage{amsfonts}
\usepackage{psfrag}
\usepackage{soul,xcolor}
\usepackage[normalem]{ulem}
\usepackage{dsfont}
\usepackage{adjustbox}
\usepackage{txfonts}

\newcommand{\ket}[1]{\vert #1 \rangle}

\begin{document}

\title{Simple analytical flux-tuned iSWAP pulses for leakage suppression}

\author{Dimitrios Georgiadis}
\affiliation{Institute for Quantum Control (PGI-8), Forschungszentrum J\"ulich, 52425 J\"ulich, Germany}
\affiliation{Institute for Theoretical Physics, University of Cologne, D-50937 Cologne, Germany}

\author{Boxi Li}
\affiliation{Institute for Quantum Control (PGI-8), Forschungszentrum J\"ulich, 52425 J\"ulich, Germany}

\author{Asier Galicia}
\affiliation{Institute for Functional Quantum System (PGI-13), Forschungszentrum J\"ulich, 52425 J\"ulich, Germany}
\affiliation{Department of Physics, RWTH Aachen University, 52074 Aachen, Germany}
\author{Rami Barends}
\affiliation{Institute for Functional Quantum System (PGI-13), Forschungszentrum J\"ulich, 52425 J\"ulich, Germany}
\affiliation{Department of Physics, RWTH Aachen University, 52074 Aachen, Germany}

\author{F. A. C\'ardenas-L\'opez}
\affiliation{Institute for Quantum Control (PGI-8), Forschungszentrum J\"ulich, 52425 J\"ulich, Germany}
\author{Felix Motzoi}
\affiliation{Institute for Quantum Control (PGI-8), Forschungszentrum J\"ulich, 52425 J\"ulich, Germany}
\affiliation{Institute for Theoretical Physics, University of Cologne, D-50937 Cologne, Germany}

\begin{abstract}
 Fast, high-fidelity two-qubit gates are a key requirement for fault-tolerant quantum computation.~Tunable coupler architectures provide a flexible approach for implementing entangling gates through flux control with large on-off ratios, but fast flux modulation can induce diabatic transitions and population leakage to non-computational states, limiting gate performance. Here we present an analytical flux control method enabling derivative removal by adiabatic gate ($\Phi$-DRAG) for suppressing leakage in flux tunable two-qubit gates. We show that $\Phi$-DRAG differs fundamentally from conventional microwave implementations and derive modified flux modulation protocols that suppress leakage below $10^{-4}$ for fast entangling gates. The method remains effective across a range of asymmetry between qubit anharmonicities and different circuit parameters, enabling high-fidelity two-qubit gates within the fifteen nanosecond range. 
\end{abstract}
\maketitle

\section{Introduction}
Superconducting quantum processors based on transmon qubits have emerged as a leading platform for scalable quantum computing due to their fast gate operations and compatibility with modern fabrication techniques \cite{Krantz2019QuantumGuideSCQubits, Kjaergaard2020SuperconductingQubits}. In recent years there has been a strong effort toward experimental realisation of quantum error correction on superconducting chips \cite{Besedin2026LatticeSurgery,Sivak2025RLQEC,Guatto2025AdaptiveQEC,rqkg-dw31}. Despite the significant progress achieved so far, the fidelity of two-qubit gates remains the primary bottleneck for the realization of fault-tolerant quantum computation, remaining only lightly below the fault-tolerance threshold.\par
Among the different approaches for realizing two-qubit interactions, tunable coupler architectures have been proposed as a versatile and scalable solution \cite{dicarlo2009demonstration,PhysRevLett.113.220502,PhysRevApplied.10.054062}. In these systems, two transmons interact through an intermediate nonlinear element whose frequency can be dynamically controlled through external magnetic flux. By modulating the coupler frequency, it is possible to engineer effective interactions between computational qubits and activate entangling gates such as CZ and iSWAP operations while maintaining very low residual coupling during idle configurations \cite{PhysRevApplied.10.054062,PhysRevApplied.20.064037,PhysRevX.11.021058}.\par
Despite significant year-by-year improvements in transmon coherence times \cite{Chang2013TiNCoherence,Place2021TantalumTransmon,Bland2025MillisecondTransmons}, faster gate operations still remain essential for achieving high gate fidelities. A major limitation of flux-controlled entangling gates, however, arises from the nonadiabatic nature of fast flux modulation. As the gate duration is reduced, the corresponding control pulses acquire broader spectral content capable of driving unwanted transitions to non-computational states. In weakly anharmonic superconducting circuits, such leakage processes constitute one of the dominant sources of coherent gate errors and fundamentally limit the achievable gate fidelity at short timescales. Although pulse shaping techniques and numerical optimal control methods have demonstrated substantial improvements \cite{Smirnov:2025cff,chen2026unlocking,PhysRevApplied.23.024059,Ding2025PulseDesign}, obtaining analytical control protocols capable of suppressing leakage while remaining experimentally feasible is still an open problem.

In microwave-driven single-qubit gates, leakage suppression is commonly achieved using the derivative removal by adiabatic gate (DRAG) framework \cite{PhysRevLett.103.110501,PhysRevA.88.062318,Jesus2025AnalyticalXGates,PRXQuantum.5.030353,Li2024CRErrorSuppression,Gao2025UltrafastSingleQubit}, where additional quadrature controls cancel off-resonant excitations through destructive interference. Extending this concept to flux-tunable architectures is non-trivial for several reasons. First, in flux-controlled systems the external drive does not directly couple to the unwanted transition matrix elements but instead modifies the effective qubit-qubit interaction indirectly through the nonlinear dependence of the coupler frequency on magnetic flux. Second, unlike microwave control, flux modulation typically provides access only to real-valued control waveforms, preventing the direct implementation of quadrature-based corrections. Finally, the effective interaction itself depends nonlinearly on the applied flux, making the relation between the physical control signal and the desired system response highly complex.\par

In this work, we develop an analytical framework for leakage suppression in flux-tunable two-qubit gates based on a generalized analytical flux-control scheme denoted $\Phi$-DRAG. Starting from a tunable-coupler transmon architecture, we derive effective control protocols that suppress leakage to non-computational states through higher-order derivative corrections applied directly to the effective interaction. We show that, despite the absence of quadrature control, leakage cancellation can still be achieved through appropriately engineered real-valued flux waveforms. In the symmetric case of equal qubit anharmonicities, we derive an analytical second-order correction that strongly suppresses leakage during fast iSWAP gates. We further extend the formalism to asymmetric systems with unequal anharmonicities by deriving fourth-order corrections capable of simultaneously cancelling multiple leakage channels.\par
Beyond the analytical derivation, we investigate the interplay between flux modulation, pulse rise time, and gate duration, demonstrating that optimized waveform shaping provides both robustness and gate-time selectivity. Our results demonstrate that leakage probabilities below $10^{-4}$ can be readily achieved for entangling gates in the tens-of-nanoseconds regime across realistic device parameters, providing a simple and experimentally accessible route toward ultra-high-fidelity flux-controlled two-qubit quantum gates.\par
\begin{figure*}[t]
    \centering
    \includegraphics[width=\textwidth]{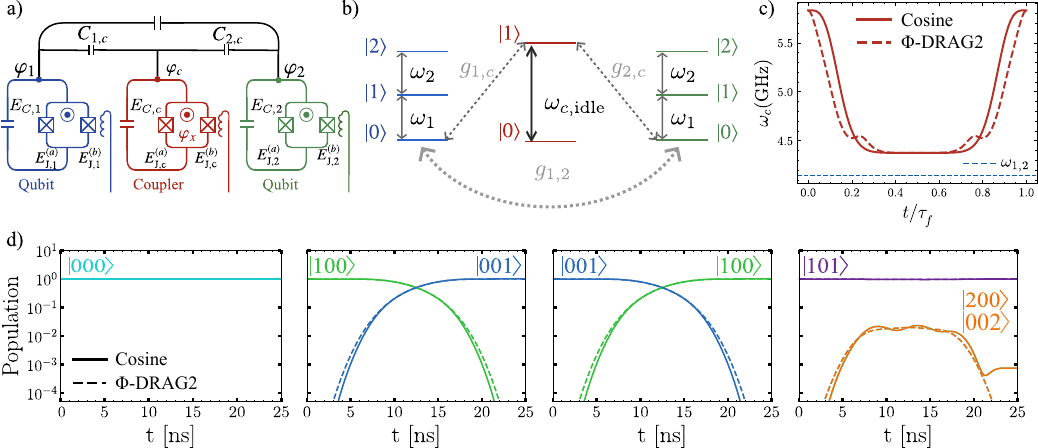}
    \caption{ \textbf{Tunable-coupler architecture and operating principle}. \textbf{a)} Circuit diagram of the tunable-coupling architecture consisting of three  tunable transmon circuits. Each transmon is characterized by its capacitance $C_{\ell}$ and Josephson energies $E_{\rm{J},\ell}^{(a)}$ and $E_{\rm{J},\ell}^{(b)}$ along a loop threaded by an external magnetic flux $\varphi_{x,\ell}$. The devices are pair-wise coupled through capacitances $C_{k,j}$. Blue represents Qb1 and Qb2, whereas the red one acts as the coupler. Each device is described by the node variable $\varphi_{\ell}$. \textbf{b)} Schematic of the working principle: by modulating the coupler flux, the effective qubit-qubit coupling $g_{\mathrm{eff}}$ is tuned from an idling point (where $g_{\mathrm{eff}} \approx 0$) to a value activating an iSWAP gate.  \textbf{c)} Implementation of the $\Phi$-DRAG correction: the corrected coupler frequency $\omega_{c}(t)$ (dashed) includes a second-derivative term in the effective coupling that suppresses leakage, compared to the initial modulated coupler frequency (solid). \textbf{d)} Time evolution of populations during a 25ns iSWAP gate, showing both the target swapping dynamics ($|100\rangle \leftrightarrow |001\rangle$) and the suppressed leakage channels ($|101\rangle \to |200\rangle,|002\rangle$) under $\Phi$-DRAG control (dashed).}
    \label{fig:circuit and pulses}
\end{figure*}
The paper is organised as follows. In Sec. II we describe the tunable-coupler architecture and derive the effective model describing flux-controlled iSWAP interactions. We then develop the analytical $\Phi$-DRAG framework, first for the symmetric case of equal anharmonicities and subsequently extend it to asymmetric systems through higher-order corrections. We further analyze the role of pulse shaping and rising-time optimization in achieving gate-time selectivity and robust leakage suppression. The performance of the proposed approach is evaluated through numerical simulations across different gate times and device imperfections. Finally, Sec. III summarizes the main results, discusses experimental considerations and outlines possible directions toward implementing $\Phi$-DRAG protocols in next-generation superconducting quantum processors.

\section{Theory}

\subsection{Model}

The superconducting circuit platform for implementing our analytical flux modulation is depicted in Fig.~\ref{fig:circuit and pulses} consisting of three transmon circuits~\cite{PhysRevA.76.042319}, each consisting of a shunted capacitor parallel connected to a superconducting interference device (SQUID) threaded by an external magnetic flux. The rightmost and leftmost transmons constitute our computational units ($Q_1$ and $Q_2$) whereas the mediator is our coupler ($C$). The qubits are directly coupled with an always-on coupling $g_{1,2}$ while they are connected to the coupler with strength $g_{\ell,c}$, respectively. Each device is modelled as an anharmonic oscillator of frequency $\omega_{\ell}$ and anharmonicity $\alpha_{\ell}$ so that the system Hamiltonian reads
\begin{eqnarray}\nonumber
\label{Hamiltonian}    \hat{\mathcal{H}}=\sum_{\ell=1,2,c}\bigg[\omega_{\ell}b_{\ell}^{\dagger}b_{\ell}+\frac{\alpha_{\ell}}{2}b_{\ell}^{\dagger}b_{\ell}^{\dagger}b_{\ell}b_{\ell}\bigg]-\sum_{k>j}g_{k,j}(b_{k}^{\dagger}-b_{k})(b_{j}^{\dagger}-b_{j}).
\end{eqnarray}
Here, we assume constant flux on the qubits so that they are fixed-frequency, while the coupler frequency is flux-tunable according to
\begin{eqnarray}
    \omega_{c}(\varphi_{\text{ext}})= \omega_{p}\big[\cos^2(\pi\varphi_{\text{ext}})+d^2\sin^2(\pi\varphi_{\text{ext}})\big]^{1/4}-E_{C}
\end{eqnarray}
where $\omega_{p}=\sqrt{8E_CE_J}$ denotes the transmon plasma frequency and $d = (E_J^{(a)} - E_J^{(b)})/(E_J^{(a)} + E_J^{(b)})$ the asymmetry of the junctions of the SQUID loop of the coupler, with $E_C$ and $E_J$ the charging and Josephson energies, respectively. Notice that for this convention $\varphi_{\rm{ext}}\in(0,0.5)$. 

In this architecture, it is possible to implement two-qubit gates between $Q_1$ and $Q_2$ mediated by the coupler in the regime when both computational units are far off-resonance with the mediator, i.e., $\omega_1,\omega_2\ll\omega_{c}(\varphi_{\text{ext}})$, $\forall \varphi_{\text{ext}}$. In such a dispersive regime, as shown previously~\cite{goerz_charting_2017}, the system dynamics is described within an effective model obtained by tracing out the coupler degree of freedom~\cite{PhysRevApplied.10.054062, PhysRev.149.491} leading to
\begin{eqnarray}\nonumber
    \label{H_eff}
    \mathcal{H}_{{\rm{eff}}}=\sum_{\ell}\bigg[\bar{\omega}_{\ell}b_{\ell}^{\dagger}b_{\ell}+\frac{\alpha_{\ell}}{2}b_{\ell}^{\dagger}b_{\ell}^{\dagger}b_{\ell}b_{\ell}\bigg]+g_{\rm{eff}}(\omega_c)(b_{1}^{\dagger}-b_{1})(b_{2}^{\dagger}-b_{2}).
\end{eqnarray}

The frequencies $\bar{\omega}_{\ell}$ represent the dressed effective frequencies of the system, while the effective coupling is given by
\begin{eqnarray}
    g_{\rm{eff}}(\omega_c)=g_{1,2}+\frac{g_{1,c}g_{2,c}}{2}\bigg[\frac{1}{\Delta_{1,c}}+\frac{1}{\Delta_{2,c}}-\frac{1}{\Sigma_{1,c}}-\frac{1}{\Sigma_{2,c}}\bigg],
    \label{g_eff}
\end{eqnarray}
where $ \Delta_{\ell,c} = \omega_{\ell} - \omega_{c}(t)$ and $ \Sigma_{\ell,c} = \omega_{\ell} + \omega_{c}(t)$ are the difference frequency (detuning) and sum frequency, respectively [see Appendix \ref{appendix} for derivation]. It is important to note that for the derivation of the effective model we have not applied the rotating wave approximation~\cite{PhysRevLett.98.013601} because the inclusion of the counter-rotating terms lead to a noticeably more accurate physical description of the systems~\cite{PhysRevApplied.10.054062}. \par

\begin{figure}[!t]
    \centering
    \includegraphics[width=1.0\linewidth]{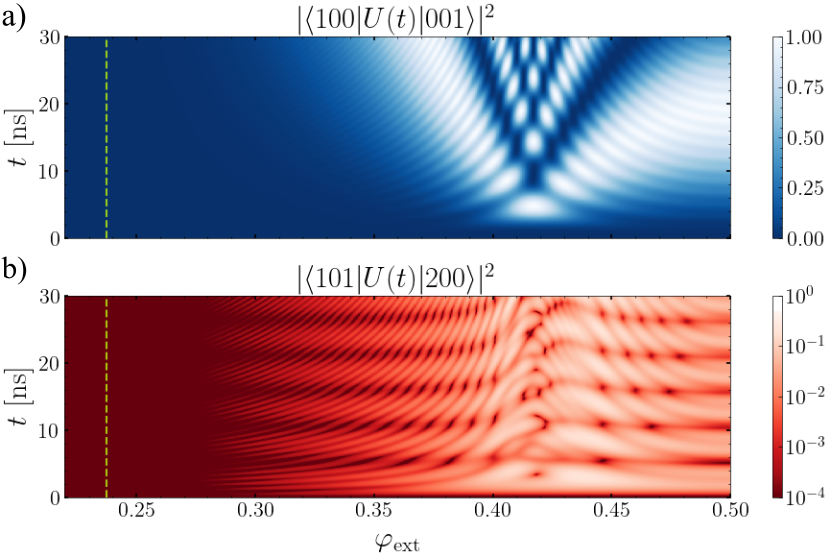}
     \caption{\textbf{Population transfer fringe patterns}. \textbf{a)} Population transfer $|\langle 100|U(t)|001\rangle|^2$ as a function of time $t$ and external coupler flux $\varphi_{\mathrm{ext}}/\varphi_0$. Chevron-like fringes indicate the iSWAP resonance condition, with faster swapping achievable at larger flux amplitudes. The green dashed line marks the idling point where $g_{\mathrm{eff}} \approx 0$. \textbf{b)} Corresponding leakage to the non-computational state, $|\langle 101|U(t)|200\rangle|^2$, on a logarithmic scale. As the flux amplitude increases (shorter gate times), leakage to $|200\rangle$ grows rapidly and becomes the dominant source of coherent error.}
     \label{fringe20}
\end{figure}

The flux-tunable architecture is a versatile platform for implementing two-qubit gates; based on the resonance conditions between the qubits it is possible to activate two different types of gates belonging to two different entanglement classes, namely the CZ and the iSWAP~\cite{PhysRevA.91.062306}. We focus on the latter aiming for very fast controlled entangling gates. We implement an iSWAP gate considering both qubits are on resonance $\omega_1=\omega_2$ and then we bias the coupler with a flux $\varphi_{ext,\rm{id}}$ where its frequency position causes  $g_{\rm{eff}}$ to vanish, i.e, the \textit{idling} configuration~\cite{PhysRevApplied.20.064037,PhysRevX.11.021058,groszkowski2011tunable}. Then we move the coupler to a maximum value $\varphi_{ext,\rm{max}}$, generating entanglement, before we modulate it back to $\varphi_{ext,\rm{id}}$ at the end of the operation $t = \tau_f$ ~\cite{PhysRevX.11.021058} thereby satisfying, 
\begin{equation}
 \int_{0}^{\tau_f} g_{\mathrm{eff}}(\varphi_{ext}(t))\,dt = \frac{\pi}{2}.
 \label{Area}
\end{equation}
Although such gates have been demonstrated experimentally in numerous works, leakage to non-computational states remains a major limitation, particularly for fast gates, as shown in Fig.~\ref{fringe20}, which shows a typical iSWAP fringe pattern. To mitigate this problem, we will implement an analytical scheme based on flux control with DRAG-inspired pulse shapes applied through the coupler.
\begin{figure*}[!t]
    \centering
    \includegraphics[width=\textwidth]{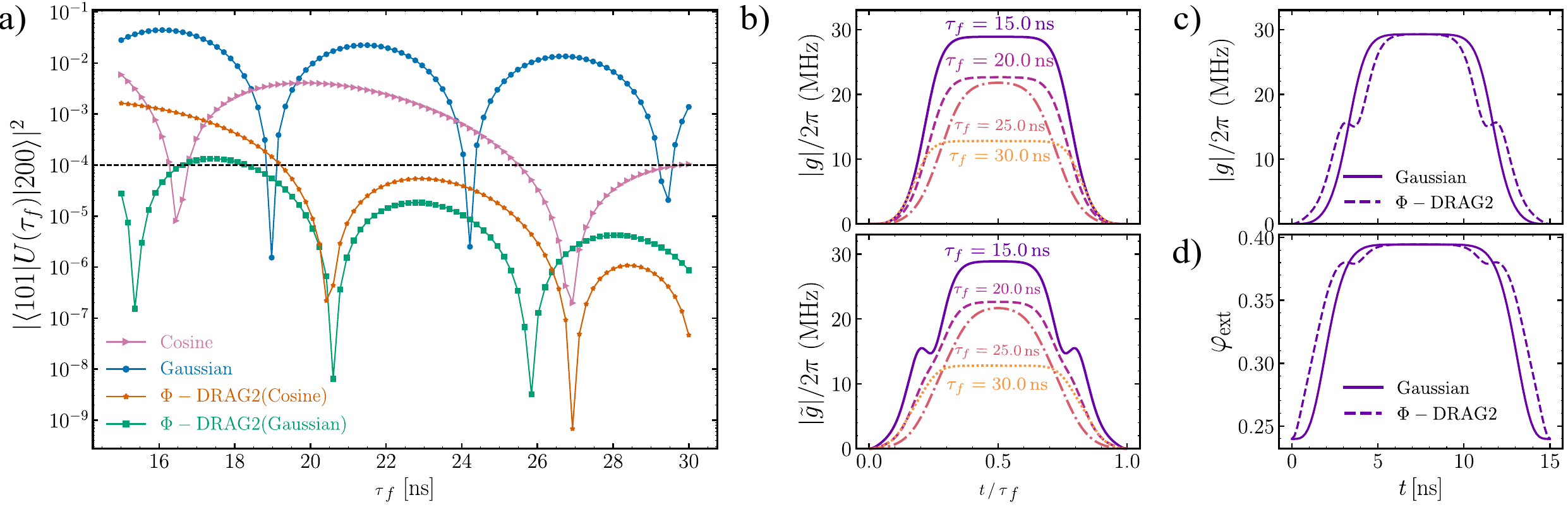}
    \caption{\textbf{Performance of $\Phi$-DRAG for the symmetric system $(\alpha_1 = \alpha_2)$}. \textbf{a)} Leakage probability to $|200\rangle$ (starting from $|101\rangle$) at the end of an iSWAP gate as a function of gate time $\tau_f$, comparing cosine-like pulses, Gaussian-like pulses Eq.~(\ref{flux_modulation_ramp}), and their $\Phi$-DRAG2-corrected counterparts Eq.~(\ref{DRAG_effcoupling}). Each data point corresponds to an iSWAP pulse whose amplitude has been optimized to satisfy the area condition Eq.~(\ref{Area}). For the Gaussian-like pulses the rising time has been selected to ensure leakage suppression below the desired threshold $10^{-4}$ (dashed line). \textbf{b)} Effective coupling $|g_{\mathrm{eff}}(t)|/2\pi$ as a function of normalized time $t/\tau_f$ for Gaussian-like pulses at four gate times $\tau_f \in \{15,20,25,30\}\,\mathrm{ns}$. \textbf{c)} Same as \textbf{b} but with the $\Phi$-DRAG2 correction applied, showing the modified coupling envelope $|\tilde{g}(t)|/2\pi$. \textbf{d)} Direct comparison of the effective coupling with (dashed) and without (solid) $\Phi$-DRAG2 correction for $\tau_f = 15\,\mathrm{ns}$, together with the corresponding physical flux waveforms $\varphi_{\mathrm{ext}}(t)$, illustrating the nonlinear mapping from effective coupling to flux.}
    \label{fig:leakage_plot and couplings}
\end{figure*}
\subsection{Control Theory for Flux}
\subsubsection{$\Phi$-DRAG}

We start by considering the truncated relevant Hilbert subspace of our system spanned by the states $\ket{200},\ket{101},\ket{002}$ written as,
\begin{eqnarray}
\hat{\mathcal{H}}_{{\rm{trun}}}(t) = 
\begin{pmatrix}
2\omega_{1} - \alpha_{1} & \sqrt{2}g_{\rm{eff}}(t) & 0 \\
\sqrt{2}g_{\rm{eff}}(t) & \omega_{1} + \omega_{2} & \sqrt{2}g_{\rm{eff}}(t) \\
0 & \sqrt{2}g_{\rm{eff}}(t) & 2\omega_{2} - \alpha_{2}
\end{pmatrix}.
\label{H_trun}
\end{eqnarray}

In the limit that the anharmonicities of the two qubits satisfy the condition $\alpha_1 = \alpha_2 = \alpha$ and the qubits being on resonance $\omega_{1} = \omega_{2}$, we can further simplify Eq.~(\ref{H_trun}) in a symmetric system as,
\begin{eqnarray}
\hat{\mathcal{H}}_{{\rm{trun,symm}}}(t) = 
\begin{pmatrix}
-\alpha & \sqrt{2}g_{\rm{eff}}(t) & 0 \\
\sqrt{2}g_{\rm{eff}}(t) & 0 & \sqrt{2}g_{\rm{eff}}(t) \\
0 & \sqrt{2}g_{\rm{eff}}(t) & -\alpha
\end{pmatrix}.
\label{H_trun_symm}
\end{eqnarray}
Our objective is to apply an analytical correction to suppress the coupling $\sqrt{2}g_{\rm{eff}}(t)$. In single qubit gates with microwave driving, it has been shown how to apply off-quadrature corrections from such amplitude crosstalk effects \cite{theis2016,h4xf-vq2l, PhysRevLett.103.110501}. In these systems, the microwave driving corresponds directly to the coupling between the undesired transitions and a generalised analytical microwave control pulse $\Omega(t)$ can be described by,
\begin{align}\label{DRAG_mw}
\Omega(t) &= \mathrm{Re}\,\Omega(t) + i\,\mathrm{Im}\,\Omega(t) \\ \nonumber
&= \Omega_0(t) + \sum_{r=1}^{n/2} a_{2r}\frac{d^{2r}}{dt^{2r}}\Omega_0(t)
+ i \sum_{r=1}^{n/2} b_{2r-1}\frac{d^{2r-1}}{dt^{2r-1}}\Omega_0(t),
\end{align}
as shown in \cite{PhysRevA.88.062318}. In our system, since the only control that we have is the shaping of the external flux through the SQUID loop of the coupler,  a direct correspondence does not exist. In particular, we do not have access to $"Y"$ control and that implies that we are constrained to only real corrections. Secondly, the effect of the flux does not translate simply into the off-diagonal elements and instead has nonlinear effects on both the couplings and the frequencies.  \par
Having that in mind, we start by modulating the waveform of the effective coupling, where we can address the leakage problem by considering only the real part of Eq.~(\ref{DRAG_mw}) and thus use the intermediate representation
\begin{align}
\tilde{g}(t)&= g_{\rm{eff}}(t) + \sum_{r=1}^{n} a_{2r}\frac{d^{2r}}{dt^{2r}}g_{\rm{eff}}(t).
\label{DRAG_effcoupling_general}
\end{align}
In order to address the excitations described by Eq.~(\ref{H_trun_symm}), since the states $\ket{200}$ and $\ket{002}$ are on resonance, it is enough to expand to the second order Eq.~(\ref{DRAG_effcoupling_general}),
\begin{align}
\tilde{g}(t)&= g_{\rm{eff}}(t) + \frac{1}{\alpha^{2}} \frac{d^{2}}{dt^{2}}g_{\rm{eff}}(t),
\label{DRAG_effcoupling}
\end{align}
where we substitute $a_{2}$ from Eq.~(\ref{DRAG_effcoupling_general}) with the detuning between the frequencies of $\ket{101}$ and $\ket{002}$, and $\ket{101}$ and $\ket{200}$, having $\Delta_{1} = - \alpha$ and $\Delta_{2} = - \alpha$ as seen in Eq~(\ref{H_trun_symm}) leading to $a_{2} = \alpha^2$. We will refer to Eq.~(\ref{DRAG_effcoupling}) as $\Phi$-DRAG2 correction.\par
We can now simulate using the circuit parameters from \cite{PhysRevX.11.021058} and compare leakage to $\ket{200}$ from Eq.~(\ref{H_eff}) with and without control corrections, as shown in Fig.~\ref{fig:leakage_plot and couplings}. Since $\ket{200}$ and $\ket{002}$ are on resonance they undergo identical dynamics. Consequently, suppressing one state automatically suppresses the other. We initially modulate the external flux as, $\varphi_{ext}(t) = \varphi_{max} + (\varphi_{id} - \varphi_{max} )\cos^{8}(\pi t/\tau_{f})$,  where $\varphi_{id}$ corresponds to the idling flux, $\varphi_{max}$ the maximum amplitude of the flux and $\tau_{f}$ the gate time. We choose this exponent in the cosine since we aim for a controllable flat-top like flux pulse shape to avoid amplitude limitations in the error. We refer to this as the cosine Ansatz.

\subsubsection{Engineering the flux pulse shape from the effective coupling}

Owing to the nonlinear relation between the effective coupling and the external flux, the experimental implementation of Eq.~(\ref{DRAG_effcoupling}) is non-trivial. In case of absence of an algorithm that inverts the coupling pulse directly to the flux pulse we can work as follows. We determine the corrected flux pulse by minimizing the cost function
\begin{equation}
\mathcal{C}(X) = \int_{0}^{\tau_f} \left[ g_{\rm{eff}}(X(t)) - \tilde{g}(t) \right]^2 \, dt,
\end{equation}
which yields a discrete representation of the corresponding new flux waveform $X(t)$. The area constraint of Eq.~(\ref{Area}) is enforced as a hard constraint during the minimization to ensure more accurate convergence. 
The resulting pulse is then approximated using a compact basis, such as Chebyshev polynomials, to obtain a smooth and experimentally feasible shape, e.g., $T_n(\cos(\theta))
=
\mathrm{Re}\bigl(\cos(n\theta)+i\sin(n\theta)\bigr)
=
\mathrm{Re}\bigl((\cos\theta+i\sin\theta)^n\bigr)$. The number of Chebyshev coefficients $n$ necessary to sample the pulse depending on the complexity of the shape which depends heavily on the speed of the gate and initial flux ansatz.

\subsection{Flux modulation and rising time control}

Since $g_{\rm{eff}}(t)$ depends on the external flux, the effect of the applied control is strongly influenced by the initial shape of the corresponding flux pulse. Despite the fact that from Fig.~\ref{fig:leakage_plot and couplings} it becomes evident that we are able to suppress sufficiently leakage to the non computational states implementing $\Phi$-DRAG in cosine flux pulses, we observe that the suppression is not monotonic with the increase of gate time. This implies that using cosine pulses and consequently controlling only the maximum amplitude of the external flux does not provide us either with adaptability or robustness in the time duration of the pulse.\par

In order to address this shortfall, we propose a Gaussian-like shape flux modulation that reads as
\begin{align}
\varphi_{ext} = \varphi_{id} + (\varphi_{max} - \varphi_{id})
\, \mathrm{erf}^{4}\left(\frac{t}{\tau_r}\right)
\, \mathrm{erf}^{4}\left(\frac{\tau_f - t}{\tau_r}\right),
\label{flux_modulation_ramp}
\end{align}
where $\tau_r$ corresponds to the rising time of the pulse. By controlling not only the maximum amplitude but also the rising time we introduce one more control parameter.
\begin{figure}[!t]
    \centering
    \includegraphics[width=1\linewidth]{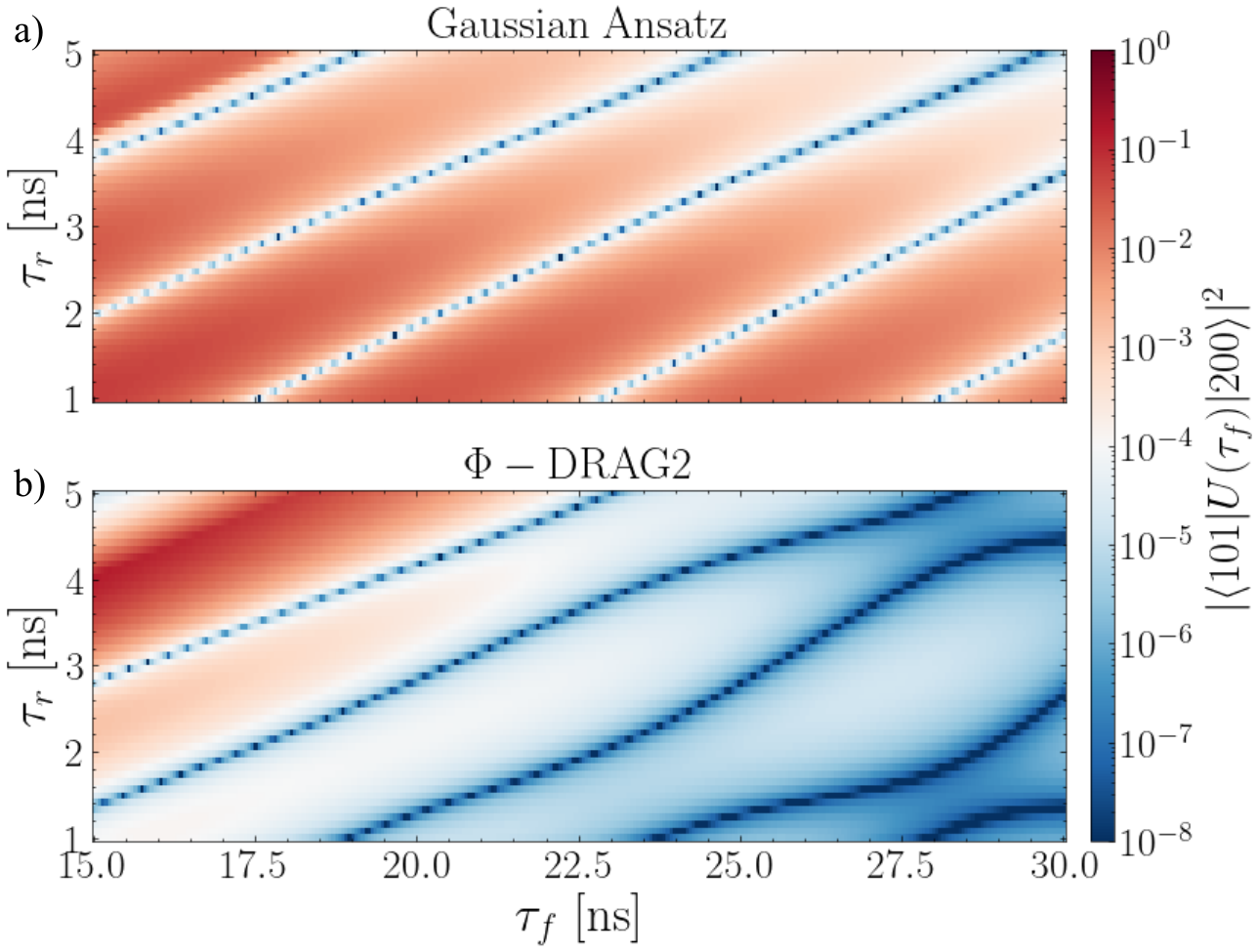}
     \caption{\textbf{Gate-time selectivity via rising time optimization for the $\Phi$-DRAG2 correction}. Leakage to $|200\rangle$ (starting from $|101\rangle$) as a function of gate time $\tau_f$ and pulse rising time $\tau_r$ for Gaussian-like pulses Eq.~(\ref{flux_modulation_ramp}). \textbf{a)} Without $\Phi$-DRAG correction: the leakage landscape shows oscillatory minima as a function of both $\tau_f$ and $\tau_r$, that require picosecond precision to achieve. \textbf{b)} With $\Phi$-DRAG2 correction Eq.~(\ref{DRAG_effcoupling}): joint optimization of rising time and gate time enables leakage below $10^{-4}$ across a broad parameter range, and the suppression minima can be steered to any desired gate time by adjusting $\tau_r$. Each point represents an area-optimized iSWAP pulse satisfying Eq.~(\ref{Area}).}
     \label{MIT_ramp_plots}
\end{figure}

The reason for that can be observed in Fig.~\ref{MIT_ramp_plots}, where by optimising the rising time of every pulse we can manipulate and move the minima of the leakage probabilities to $\ket{200}$, thereby enabling gate time selectivity.\par

Furthermore, applying the correction given in Eq.~(\ref{DRAG_effcoupling}) while using the above mentioned flux modulation demonstrates that very fast gates can be realized while maintaining adaptability and adding robustness in terms of preserving leakage below $10^{-4}$, as seen in Fig.~\ref{fig:leakage_plot and couplings} for gate times below $19\text{ns}$. Moreover, depending on the precision of current experimental instruments, improvements in leakage suppression with an average of about three orders of magnitude and sometimes exceeding six orders of magnitude can be achieved, rendering the leakage effectively negligible Fig.~\ref{MIT_ramp_plots}.

Finally we notice that a decrease in rising time is necessary the faster the gate. This can be justified by amplitude restriction, considering that, apart from satisfying the condition to maintain the area of the pulse equal to $\frac{\pi}{2}$, we are also restricted by the dispersive approximation i.e. ($\omega_{c_{min}} - \omega_{qubit}) \gg 0 $ \cite{goerz_charting_2017}.

\subsection{Beyond the symmetrical system}

While Eq.~(\ref{DRAG_effcoupling}), as illustrated in Fig.~\ref{MIT_ramp_plots}, provides an effective solution for leakage suppression, it is important to emphasize the assumptions underlying its derivation. In particular, the result was obtained under the conditions that the qubits are operated on resonance and possess identical anharmonicities. Although tuning the qubits into resonance via flux control is experimentally straightforward, the assumption of obtaining precisely equal anharmonicities is generally not guaranteed in practice, as fabrication-induced variations typically lead to device-dependent anharmonicities. Such a scenario can be described more generally if in Eq.~(\ref{H_trun}) we only assume the resonant condition $\omega_{1} = \omega_{2}$, leading to
\begin{eqnarray}
\hat{\mathcal{H}}_{{\rm{trun}}}(t) = 
\begin{pmatrix}
-\alpha_1 & \sqrt{2}g_{\rm{eff}}(t) & 0 \\
\sqrt{2}g_{\rm{eff}}(t) & 0 & \sqrt{2}g_{\rm{eff}}(t) \\
0 & \sqrt{2}g_{\rm{eff}}(t) & -\alpha_2
\end{pmatrix}.
\label{H_trun_4th}
\end{eqnarray}
We immediately observe that now we have two disjoint transitions that are not on resonance and we want to suppress simultaneously.\par 
In order to achieve that we need to extend Eq.~(\ref{DRAG_effcoupling_general}) to the fourth order where $a_{2}$ and $a_{4}$ are derived as,
\begin{align}
a_2 &= \frac{1}{\alpha_1^2} + \frac{1}{\alpha_2^2}, \qquad a_4 = \frac{1}{\alpha_1^2 \alpha_2^2},
\end{align}
leading to the equation that we will refer as $\Phi$-DRAG4,

\begin{align}
\tilde{g}(t)&= g_{\rm{eff}}(t) + \left(\frac{1}{\alpha_{1}^{2}} + \frac{1}{\alpha_{2}^{2}}\right)\frac{d^{2}}{dt^{2}}g_{\rm{eff}}(t) + \left(\frac{1}{\alpha_{1}^{2}\alpha_{2}^{2}}\right)\frac{d^{4}}{dt^{4}}g_{\rm{eff}}(t). 
\label{4th_DRAG_effcoupling}
\end{align}
\begin{figure}[!b]
    \centering
    \includegraphics[width=1.0\linewidth]{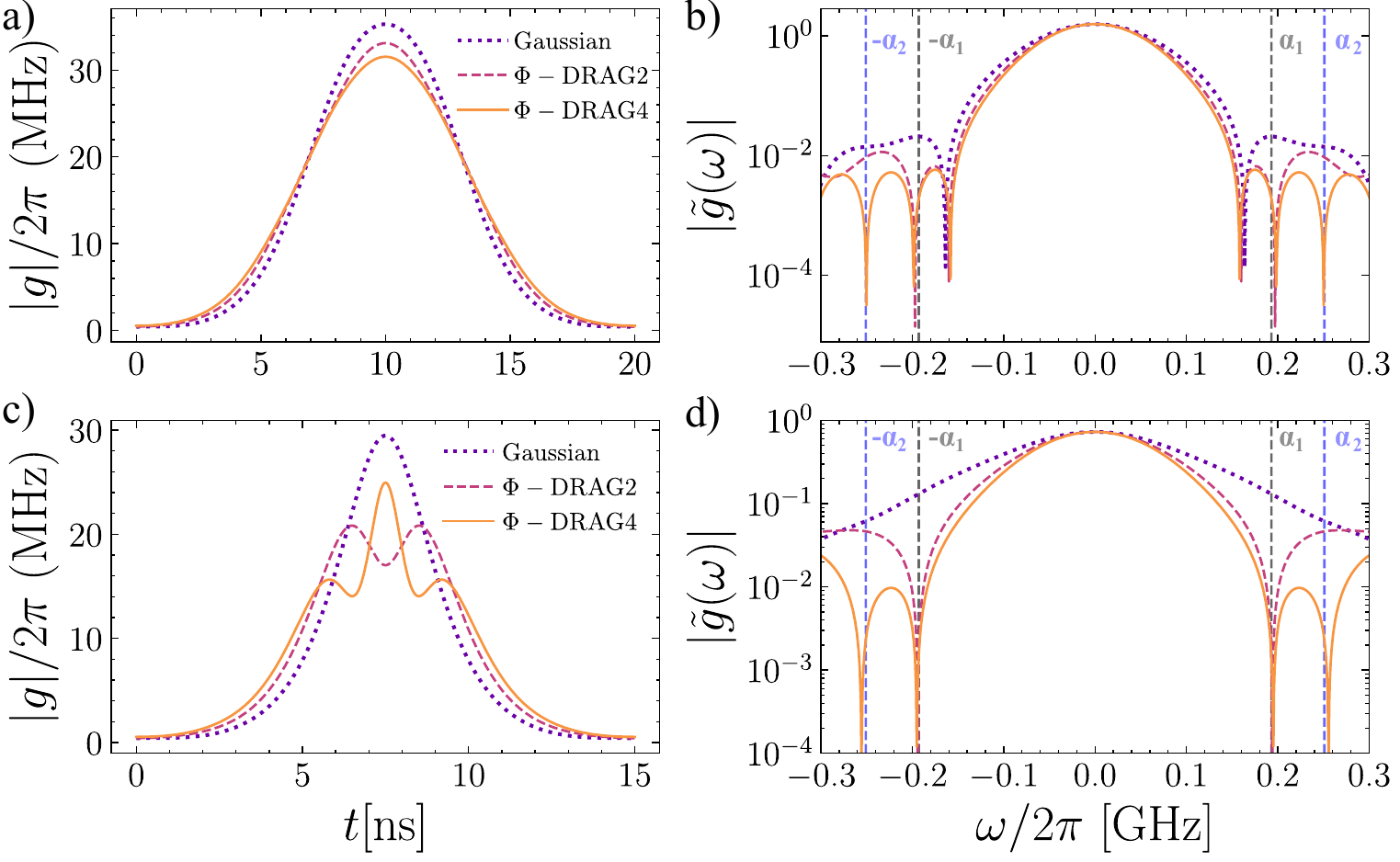}
     \caption{Pulse shapes and Fourier spectra for second- and fourth-order $\Phi$-DRAG corrections. \textbf{a)} Effective coupling $g_{\mathrm{eff}}(t)/2\pi$ as a function of time (left column) and its Fourier spectrum $|\tilde{g}(\omega)|/2\pi$ (right column), shown for both the bare (initial) corresponding to Eq.~(\ref{flux_modulation_ramp}), $\Phi$-DRAG2, and $\Phi$-DRAG4 waveforms. \textbf{a,b)} $\tau_f = 20\,\mathrm{ns}$ gate: the $\Phi$-DRAG2 correction creates spectral nulls at $\pm \alpha_1/2\pi$ (the transition frequency driving leakage), while $\Phi$-DRAG4 additionally addresses the asymmetric case with nulls at $\pm \alpha_1/2\pi$ and $\pm \alpha_2/2\pi$. \textbf{c,d)} $\tau_f = 15\,\mathrm{ns}$ gate: the broader spectral content required for faster gates leads to more complex corrected waveforms, making the $\Phi$-DRAG4 pulse more susceptible to AWG bandwidth limitations and waveform distortions.}
     \label{alphas_fft}
\end{figure}
\begin{figure*}[t!]
    \centering
    \includegraphics[width=\textwidth]{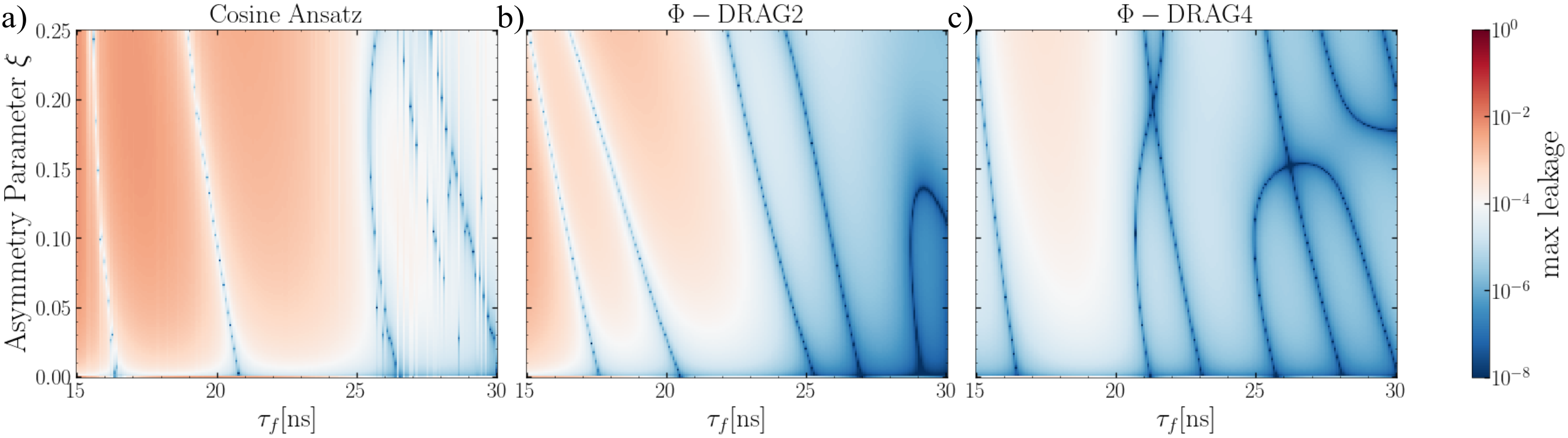}
    \caption{\textbf{Robustness of $\Phi$-DRAG against anharmonicity asymmetry}. Maximum leakage between $|200\rangle$ and $|002\rangle$ (starting from $|101\rangle$) as a function of gate time $\tau_f$ and anharmonicity asymmetry parameter $\xi$, defined by $\alpha_1 = (1+\xi)\alpha_2$. \textbf{a)} Cosine ansatz iSWAP pulses without correction: leakage exceeds $10^{-3}$ across the majority of parameter space for fast gates. \textbf{b)}  With second-order $\Phi$-DRAG correction ($\Phi$-DRAG2): leakage is suppressed below $10^{-4}$ for asymmetries up to $5\%$ for gate times above $17\,\mathrm{ns}$, and remains below $10^{-4}$ showing robustness for slower gates no matter the asymmetry. \textbf{c)} With fourth-order $\Phi$-DRAG correction ($\Phi$-DRAG4): leakage suppression extends to larger asymmetries even for very fast gates, demonstrating robustness against realistic fabrication-induced variations in qubit anharmonicities.}
    \label{fig:robustness_alphas}
\end{figure*}

In the regime \( \tilde{g}(t) \ll \alpha \), the fourth-order solution is expected to provide sufficient suppression of leakage to the \(\ket{200}\) and \(\ket{002}\) states, even for relatively fast gate operations. As seen in Fig.~\ref{alphas_fft} the second order correction creates a hole in the Fourier spectrum for the transitions corresponding to $\pm \alpha$ while the fourth order addresses also the ones if $\alpha_{1} \neq \alpha_{2}$. However, an important consideration arises from the fact that reducing the gate duration typically requires increasing the flux amplitude. The inclusion of higher-order corrections further enhances this amplitude, which may ultimately compromise the validity of the dispersive approximation made in Eq.~(\ref{g_eff}). One way to moderate this effect is by implementing the flux modulation of Eq.~(\ref{flux_modulation_ramp}), where by adjusting the rising time of the pulse we can control better the shape and thus the maximum amplitude as seen in Fig~\ref{alphas_fft}(a).\par

In Fig.~\ref{fig:robustness_alphas}, we can observe that the second order solution still holds for up to $10\%$ asymmetry between the anharmonicities for fast gates below $20\text{ns}$ while for slower gates even up to $25\%$ asymmetry population of \(\ket{200}\) remains below $10^{-4}$.

Despite the fact that the fourth order solution seems to outperform second order for fast gates, we observe as seen in Fig.~\ref{alphas_fft}(b) that the shape of the pulse becomes more complicated the shorter the pulse and thus becomes more sensitive to distortion errors \cite{1qhb-r4fb,rol2020cryoscope,barends2014surfacecode} as well as to bandwidth and sampling limitations of experimental tools like AWGs \cite{motzoi2011optimal,PhysRevApplied.12.044054,landi2012characterization}. This leads us to the conclusion that with further improvement of experimental instruments and transfer function matching \cite{singh2023compensating}, we can aim for better leakage suppression even at the quantum speed limit of the gate for every flux tunable architecture based chip.\par

\section{DISCUSSION and conclusion}
In this work we demonstrate how leakage suppression in a fast flux-tunable iSWAP gate can be readily achieved using simple analytical corrections inspired by DRAG techniques while remaining compatible with current experimental feasible flux control tools. The intricacies of the non-linear relation between flux driving and coupling between undesired transitions fundamentally differs from microwave control and imposes additional constraints.\par
Within the symmetric limit of equal qubit anharmonicities, we observe that the second-order correction ($\Phi$-DRAG2) already suppresses population transfer to non-computational states below $10^{-4}$ for gate times above the 15 nanoseconds regime for realistic circuit parameters. Extending the method to asymmetric systems requires fourth order corrections ($\Phi$-DRAG4) to simultaneously suppress leakage reliably to both qubits for fast gates.
Nevertheless we find that, with additional rising time control of the flux-pulse not only, we not only enable gate-time selectivity but also the second-order correction remains surprisingly robust for moderate anharmonicity mismatch for gates above $20\text{ns}$, suggesting tolerance against realistic fabrication variations.\par
From an experimental perspective, implementation of higher-order $\Phi$-DRAG corrections may be limited by waveform distortions and finite AWG bandwidth and sampling rates. However, ongoing improvements in control electronics suggest that increasingly complex pulse shapes may become accessible, potentially enabling leakage suppression close to the intrinsic speed limits of tunable-coupler architectures. 

\section{Acknowledgements}

We are grateful to Manuel Guatto, Jos\'e Jesus, Aneesh Kamat, Shahrukh
Chishti, Nermine Chaabani and Yuan Gao for valuable discussions and insightful feedback. This work was supported by the German Federal Ministry of Education and Research (BMBF) through the QSolid project (Grant No. 13N16149), the ML4Q2 Cluster of Excellence (EXC 2004/2 – 390534769), and the European Union's  Horizon Europe programme through the OpenSuperQPlus100 project (Grant No. 101113946, HORIZON-CL4-2022-QUANTUM-01-SGA) through the QCFD project (Grant No. 101080085, HORIZON-CL4-2021-DIGITAL-EMERGING-02-10).

\appendix{}

\section{SWT and counter rotating terms}
\label{appendix}
The absence of leakage and phase errors towards excited states of the coupler relies on assuming that at any time the architecture be dispersively coupled, i.e., $\{\omega_1,\omega_2\}\ll \omega_{c}$. Under this condition, we  \textit{dynamically factorize} Qb1 and Qb2 from the coupler and an effective two-qubit interaction arises. The effective Hamiltonian is obtained applying the Schrieffer-Wolff-Transformation~\cite{PhysRev.149.491}. 
\begin{eqnarray}\nonumber
    \hat{S}(t)=\sum_{n}\xi_{n}(b_{n}^{\dagger}b_{c}-b_{n}b_{c}^{\dagger}),
\end{eqnarray}
where the index $n$ runs for $\{\text{Qb1,Qb2}\}$. At second order, we obtain that the effective Hamiltonian reads
\begin{eqnarray}\nonumber
    \mathcal{H}_{{\rm{eff}}}=\mathcal{H}+[\hat{S},\mathcal{H}]+\frac{1}{2!}[\hat{S},[\hat{S},\mathcal{H}]]
\end{eqnarray}
To cancel the off-diagonal interaction, we demand that the generator satisfy the following relation: $\mathcal{H}_{I}+[\hat{S},\mathcal{H}_{0}]=0$, which leads to the following effective Hamiltonian
\begin{eqnarray}
    \mathcal{H}_{{\rm{eff}}}=\mathcal{H}_{0}+\frac{1}{2}[\hat{S},\mathcal{H}_{I}],
\end{eqnarray}
In this case, the SWT generator is given by
\begin{eqnarray}\nonumber
    \hat{S}=\sum_{n}\bigg[\frac{g_{n,c}}{\Delta_{n,c}}(b_{n}^{\dagger}b_{c}-b_{n}b_{c}^{\dagger}) - \frac{g_{n,c}}{\Sigma_{n,c}}(b_{n}^{\dagger}b_{c}^{\dagger}-b_{n}b_{c})\bigg].
\end{eqnarray}
 Note that the structure of the generator include two terms because we need to simultaneously diagonalize the rotating and the counter rotating term of the Hamiltonian. That lead to,
\begin{eqnarray}\nonumber
    \mathcal{H}_{{\rm{eff}}}&=&\sum_{\ell}\bigg[\bar{\omega}_{\ell}b_{\ell}^{\dagger}b_{\ell}+\frac{\eta_{\ell}}{2}b_{\ell}^{\dagger}b_{\ell}^{\dagger}b_{\ell}b_{\ell}\bigg]\\
    &+&g_{\rm{eff}}(\omega_c)(b_{1}^{\dagger}-b_{1})(b_{2}^{\dagger}-b_{2}).
\end{eqnarray}
The effective frequencies and coupling for QB1 and QB2 and the coupler are given by 
\begin{subequations}
    \begin{eqnarray}
    \bar{\omega}_{n}&=&\omega_{n}-\frac{4g_{n,c}^2}{\Sigma_{n,c}}+\frac{2g_{n,c}^2}{\Delta_{n,c}}-\frac{2g_{k,c}^2}{\Sigma_{k,c}},\quad \\
    \bar{\omega}_{c}&=&\omega_{c}-\bigg[\frac{4g_{1,c}^2}{\Sigma_{1,c}}-\frac{2g_{2,c}^2}{\Delta_{2,c}}\bigg],\\
    g_{\rm{eff}}(\omega_c)&=&g_{1,2}+\frac{g_{1,c}g_{2,c}}{2}\bigg[\frac{1}{\Delta_{1,c}}+\frac{1}{\Delta_{2,c}}-\frac{1}{\Sigma_{1,c}}-\frac{1}{\Sigma_{2,c}}\bigg].~~~~~~
    \label{gef}
\end{eqnarray}
\end{subequations}
\begin{figure}[!h]
    \centering
    \includegraphics[width=0.8\linewidth]{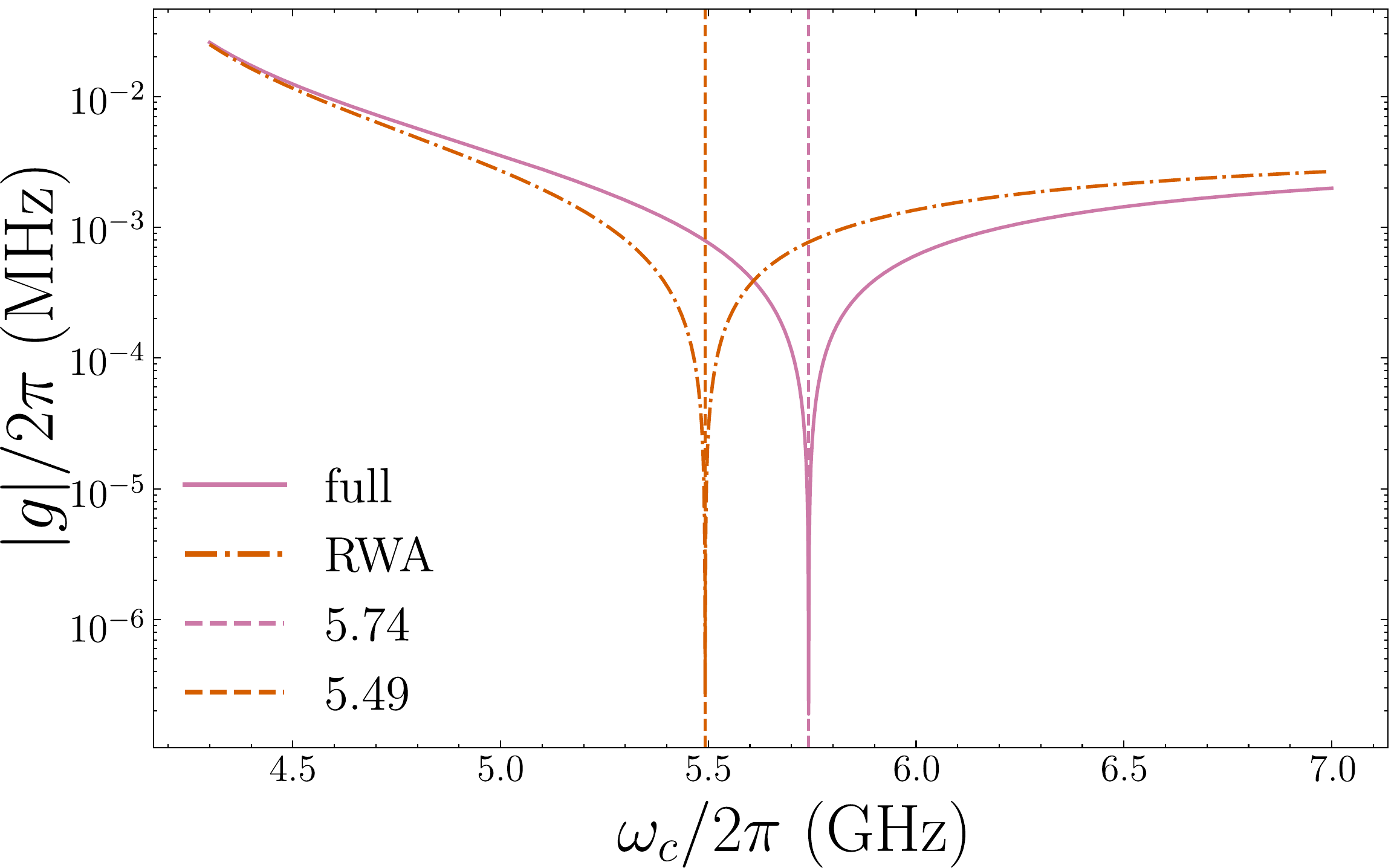}
     \caption{Importance of counter-rotating terms in the effective coupling. Effective coupling $g_{\mathrm{eff}}/2\pi$ as a function of coupler frequency $\omega_c/2\pi$, comparing the full Schrieffer--Wolff result including counter-rotating terms (Eq.~(\ref{gef}), solid lines) with the rotating wave approximation (RWA, dash-dotted lines). The two predictions differ by approximately $250\,\mathrm{MHz}$ at the idling point where $g_{\mathrm{eff}}=0$: the full expression predicts $\omega_c/2\pi \approx 5.74\,\mathrm{GHz}$ while the RWA gives $\omega_c/2\pi \approx 5.49\,\mathrm{GHz}$.}
     \label{geff_RWA}
\end{figure}
The importance of counter rotating terms \cite{PhysRevApplied.10.054062} $\Sigma_{\ell,c}$ can be observed in Fig.~\ref{geff_RWA} where we notice a mismatch of $250\text{MHz}$ at the idling point if we use the RWA.


\section{Derivation of $\Phi$-DRAG coefficients for symmetrical and asymmetrical system}
We start from the equation,
\begin{align}
\tilde{g(t)}&= g_{\rm{eff}}(t) + \sum_{r=1}^{n} a_{2r}\frac{d^{2r}}{dt^{2r}}g_{\rm{eff}}(t),
\end{align}
where assuming that the states \(\ket{200}\) and \(\ket{002}\) are energetically on resonance we can think of having $n=1$ meaning only one transition to suppress. In order to calculate the parameter $a_{2}$ we follow Eq.(3.8) in \cite{PhysRevA.88.062318} where in our case we keep only the first part of the equation,
\begin{align}
1 + \sum_{r=1}^{n}(-1)^{r}(\Delta)^{2r}a_{2r}= 0.
\end{align}
We have that from Eq.~(\ref{H_trun_symm}), $\Delta = \alpha$ and thus $a_{2} = \frac{1}{\alpha^2}$. In the same way we can calculate $a_{2}$ and $a_{4}$ in the case that $\alpha_{1} \neq \alpha_{2}$. The equations that we need to solve read,
\begin{subequations}
    \begin{eqnarray}
        &&1 - (\Delta_{1})^{2}a_{2} + (\Delta_{1})^{4}a_{4}= 0,\\
        &&1 - (\Delta_{2})^{2}a_{2} + (\Delta_{2})^{4}a_{4}= 0,
    \end{eqnarray}
\end{subequations}
having $\Delta_{1} = \alpha_{1}$ and $\Delta_{2} = \alpha_{2}$ leading to,
\begin{align}
a_2 &= \frac{1}{\alpha_1^2} + \frac{1}{\alpha_2^2}, \qquad a_4 = \frac{1}{\alpha_1^2 \alpha_2^2}.
\end{align}

\section{Robustness to sampling rate}

\begin{figure}[!h]
    \centering
    \includegraphics[width=1\linewidth]{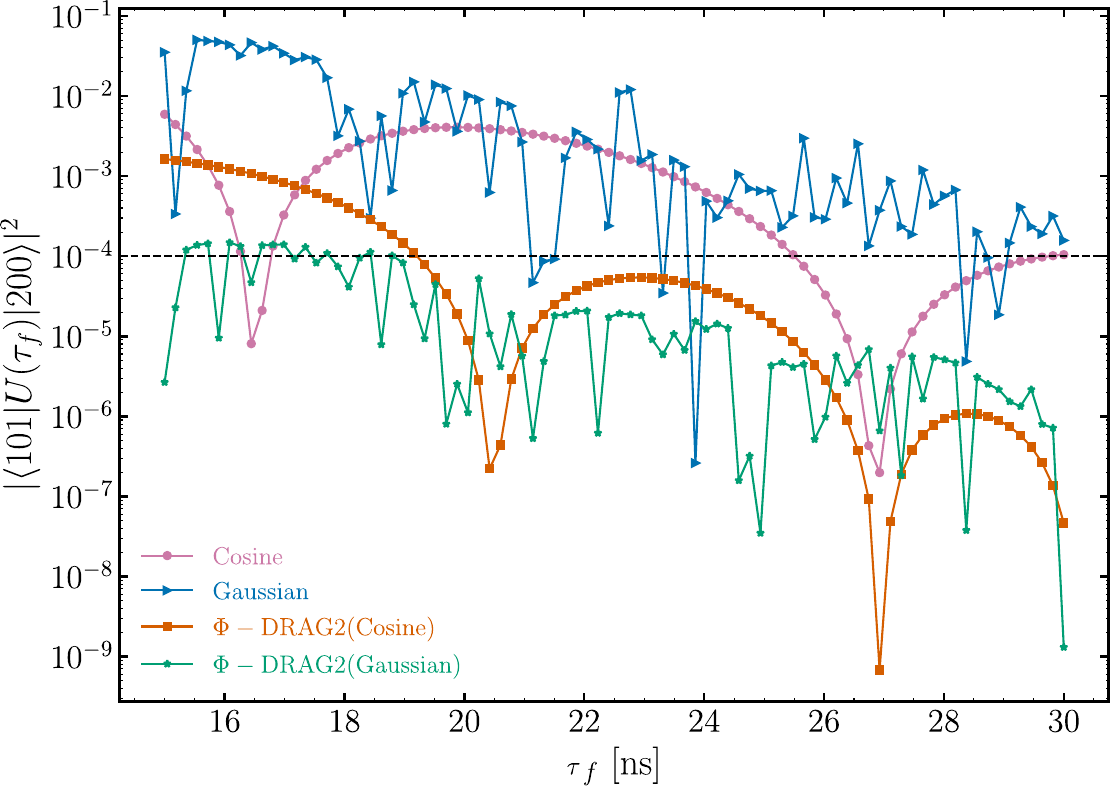}
     \caption{Performance of $\Phi$-DRAG for the symmetric system $(\alpha_1 = \alpha_2)$. Leakage probability to $|200\rangle$ (starting from $|101\rangle$) at the end of an iSWAP gate as a function of gate time $\tau_f$, comparing cosine-like pulses, Gaussian-like pulses Eq.~(\ref{flux_modulation_ramp}), and their $\Phi$-DRAG2-corrected counterparts Eq.~(\ref{DRAG_effcoupling}). Each data point corresponds to an iSWAP pulse whose amplitude has been optimized to satisfy the area condition Eq.~(\ref{Area}). For the Gaussian-like pulses the rising time has been optimized under 0.5~ns sampling rate limitations.}
     \label{rob_erf}
\end{figure}
The results presented in Fig.~\ref{fig:leakage_plot and couplings}(a) were obtained using an optimized Gaussian-like flux pulse with a rise time of $1.5$~ns, as shown in Fig.~\ref{MIT_ramp_plots}. Although such pulse parameters are readily accessible in numerical simulations, practical implementations are constrained by the finite temporal resolution of arbitrary waveform generators (AWGs). These limitations may require larger rise times and introduce waveform discretization effects that could potentially degrade the performance of the leakage-suppression protocol.\par

To evaluate the robustness of $\Phi$-DRAG2 under realistic hardware conditions, we simulate both the corrected and uncorrected Gaussian pulse ansätze using a sampling interval of $0.5$~ns and a range of larger rise times. The objective is not to further optimize the pulse, but rather to determine whether the leakage suppression obtained for the ideal waveform survives under experimentally relevant constraints.\par
As shown in Fig.~\ref{rob_erf}, the performance of $\Phi$-DRAG2 remains remarkably robust. Despite the reduced temporal resolution and the use of longer rise times, leakage into non-computational states remains below the target threshold throughout the investigated parameter range. These results indicate that the correction scheme does not rely on finely tuned pulse edges and can therefore be implemented using currently available control hardware while retaining its effectiveness in the ultra-fast gate regime.

\appendix


\begin{thebibliography}{39}%
\makeatletter
\providecommand \@ifxundefined [1]{%
 \@ifx{#1\undefined}
}%
\providecommand \@ifnum [1]{%
 \ifnum #1\expandafter \@firstoftwo
 \else \expandafter \@secondoftwo
 \fi
}%
\providecommand \@ifx [1]{%
 \ifx #1\expandafter \@firstoftwo
 \else \expandafter \@secondoftwo
 \fi
}%
\providecommand \natexlab [1]{#1}%
\providecommand \enquote  [1]{``#1''}%
\providecommand \bibnamefont  [1]{#1}%
\providecommand \bibfnamefont [1]{#1}%
\providecommand \citenamefont [1]{#1}%
\providecommand \href@noop [0]{\@secondoftwo}%
\providecommand \href [0]{\begingroup \@sanitize@url \@href}%
\providecommand \@href[1]{\@@startlink{#1}\@@href}%
\providecommand \@@href[1]{\endgroup#1\@@endlink}%
\providecommand \@sanitize@url [0]{\catcode `\\12\catcode `\$12\catcode
  `\&12\catcode `\#12\catcode `\^12\catcode `\_12\catcode `\%12\relax}%
\providecommand \@@startlink[1]{}%
\providecommand \@@endlink[0]{}%
\providecommand \url  [0]{\begingroup\@sanitize@url \@url }%
\providecommand \@url [1]{\endgroup\@href {#1}{\urlprefix }}%
\providecommand \urlprefix  [0]{URL }%
\providecommand \Eprint [0]{\href }%
\providecommand \doibase [0]{https://doi.org/}%
\providecommand \selectlanguage [0]{\@gobble}%
\providecommand \bibinfo  [0]{\@secondoftwo}%
\providecommand \bibfield  [0]{\@secondoftwo}%
\providecommand \translation [1]{[#1]}%
\providecommand \BibitemOpen [0]{}%
\providecommand \bibitemStop [0]{}%
\providecommand \bibitemNoStop [0]{.\EOS\space}%
\providecommand \EOS [0]{\spacefactor3000\relax}%
\providecommand \BibitemShut  [1]{\csname bibitem#1\endcsname}%
\let\auto@bib@innerbib\@empty
\bibitem [{\citenamefont {Krantz}\ \emph {et~al.}(2019)\citenamefont {Krantz},
  \citenamefont {Kjaergaard}, \citenamefont {Yan}, \citenamefont {Orlando},
  \citenamefont {Gustavsson},\ and\ \citenamefont
  {Oliver}}]{Krantz2019QuantumGuideSCQubits}%
  \BibitemOpen
  \bibfield  {author} {\bibinfo {author} {\bibfnamefont {P.}~\bibnamefont
  {Krantz}}, \bibinfo {author} {\bibfnamefont {M.}~\bibnamefont {Kjaergaard}},
  \bibinfo {author} {\bibfnamefont {F.}~\bibnamefont {Yan}}, \bibinfo {author}
  {\bibfnamefont {T.~P.}\ \bibnamefont {Orlando}}, \bibinfo {author}
  {\bibfnamefont {S.}~\bibnamefont {Gustavsson}},\ and\ \bibinfo {author}
  {\bibfnamefont {W.~D.}\ \bibnamefont {Oliver}},\ }\bibfield  {title}
  {\bibinfo {title} {A quantum engineer's guide to superconducting qubits},\
  }\href {https://doi.org/10.1063/1.5089550} {\bibfield  {journal} {\bibinfo
  {journal} {Applied Physics Reviews}\ }\textbf {\bibinfo {volume} {6}},\
  \bibinfo {pages} {021318} (\bibinfo {year} {2019})}\BibitemShut {NoStop}%
\bibitem [{\citenamefont {Kjaergaard}\ \emph {et~al.}(2020)\citenamefont
  {Kjaergaard}, \citenamefont {Schwartz}, \citenamefont {Braum{\"u}ller},
  \citenamefont {Krantz}, \citenamefont {Wang}, \citenamefont {Gustavsson},\
  and\ \citenamefont {Oliver}}]{Kjaergaard2020SuperconductingQubits}%
  \BibitemOpen
  \bibfield  {author} {\bibinfo {author} {\bibfnamefont {M.}~\bibnamefont
  {Kjaergaard}}, \bibinfo {author} {\bibfnamefont {M.~E.}\ \bibnamefont
  {Schwartz}}, \bibinfo {author} {\bibfnamefont {J.}~\bibnamefont
  {Braum{\"u}ller}}, \bibinfo {author} {\bibfnamefont {P.}~\bibnamefont
  {Krantz}}, \bibinfo {author} {\bibfnamefont {J.~I.-J.}\ \bibnamefont {Wang}},
  \bibinfo {author} {\bibfnamefont {S.}~\bibnamefont {Gustavsson}},\ and\
  \bibinfo {author} {\bibfnamefont {W.~D.}\ \bibnamefont {Oliver}},\ }\bibfield
   {title} {\bibinfo {title} {Superconducting qubits: Current state of play},\
  }\href {https://doi.org/10.1146/annurev-conmatphys-031119-050605} {\bibfield
  {journal} {\bibinfo  {journal} {Annual Review of Condensed Matter Physics}\
  }\textbf {\bibinfo {volume} {11}},\ \bibinfo {pages} {369} (\bibinfo {year}
  {2020})}\BibitemShut {NoStop}%
\bibitem [{\citenamefont {Besedin}\ \emph {et~al.}(2026)\citenamefont
  {Besedin}, \citenamefont {Kerschbaum}, \citenamefont {Knoll},\ and\
  \citenamefont {et~al.}}]{Besedin2026LatticeSurgery}%
  \BibitemOpen
  \bibfield  {author} {\bibinfo {author} {\bibfnamefont {I.}~\bibnamefont
  {Besedin}}, \bibinfo {author} {\bibfnamefont {M.}~\bibnamefont {Kerschbaum}},
  \bibinfo {author} {\bibfnamefont {J.}~\bibnamefont {Knoll}},\ and\ \bibinfo
  {author} {\bibnamefont {et~al.}},\ }\bibfield  {title} {\bibinfo {title}
  {Lattice surgery realized on two distance-three repetition codes with
  superconducting qubits},\ }\href {https://doi.org/10.1038/s41567-025-03090-6}
  {\bibfield  {journal} {\bibinfo  {journal} {Nature Physics}\ }\textbf
  {\bibinfo {volume} {22}},\ \bibinfo {pages} {189} (\bibinfo {year}
  {2026})}\BibitemShut {NoStop}%
\bibitem [{\citenamefont {Sivak}\ \emph {et~al.}(2025)\citenamefont {Sivak},
  \citenamefont {Morvan}, \citenamefont {Broughton}, \citenamefont {Neeley},
  \citenamefont {Eickbusch}, \citenamefont {Abanin} \emph
  {et~al.}}]{Sivak2025RLQEC}%
  \BibitemOpen
  \bibfield  {author} {\bibinfo {author} {\bibfnamefont {V.}~\bibnamefont
  {Sivak}}, \bibinfo {author} {\bibfnamefont {A.}~\bibnamefont {Morvan}},
  \bibinfo {author} {\bibfnamefont {M.}~\bibnamefont {Broughton}}, \bibinfo
  {author} {\bibfnamefont {M.}~\bibnamefont {Neeley}}, \bibinfo {author}
  {\bibfnamefont {A.}~\bibnamefont {Eickbusch}}, \bibinfo {author}
  {\bibfnamefont {D.}~\bibnamefont {Abanin}}, \emph {et~al.},\ }\href@noop {}
  {\bibinfo {title} {Reinforcement learning control of quantum error
  correction}} (\bibinfo {year} {2025}),\ \Eprint
  {https://arxiv.org/abs/2511.08493} {arXiv:2511.08493 [quant-ph]} \BibitemShut
  {NoStop}%
\bibitem [{\citenamefont {Guatto}\ \emph {et~al.}(2025)\citenamefont {Guatto},
  \citenamefont {Preti}, \citenamefont {Schilling}, \citenamefont {Calarco},
  \citenamefont {C{\'a}rdenas-L{\'o}pez},\ and\ \citenamefont
  {Motzoi}}]{Guatto2025AdaptiveQEC}%
  \BibitemOpen
  \bibfield  {author} {\bibinfo {author} {\bibfnamefont {M.}~\bibnamefont
  {Guatto}}, \bibinfo {author} {\bibfnamefont {F.}~\bibnamefont {Preti}},
  \bibinfo {author} {\bibfnamefont {M.}~\bibnamefont {Schilling}}, \bibinfo
  {author} {\bibfnamefont {T.}~\bibnamefont {Calarco}}, \bibinfo {author}
  {\bibfnamefont {F.~A.}\ \bibnamefont {C{\'a}rdenas-L{\'o}pez}},\ and\
  \bibinfo {author} {\bibfnamefont {F.}~\bibnamefont {Motzoi}},\ }\href@noop {}
  {\bibinfo {title} {Real-time adaptive quantum error correction by model-free
  multi-agent learning}} (\bibinfo {year} {2025}),\ \Eprint
  {https://arxiv.org/abs/2509.03974} {arXiv:2509.03974 [quant-ph]} \BibitemShut
  {NoStop}%
\bibitem [{\citenamefont {He}\ \emph {et~al.}(2025)\citenamefont {He},
  \citenamefont {Lin},\ and\ \citenamefont {et~al.}}]{rqkg-dw31}%
  \BibitemOpen
  \bibfield  {author} {\bibinfo {author} {\bibfnamefont {T.}~\bibnamefont
  {He}}, \bibinfo {author} {\bibfnamefont {W.}~\bibnamefont {Lin}},\ and\
  \bibinfo {author} {\bibnamefont {et~al.}},\ }\bibfield  {title} {\bibinfo
  {title} {Experimental quantum error correction below the surface code
  threshold via all-microwave leakage suppression},\ }\href
  {https://doi.org/10.1103/rqkg-dw31} {\bibfield  {journal} {\bibinfo
  {journal} {Phys. Rev. Lett.}\ }\textbf {\bibinfo {volume} {135}},\ \bibinfo
  {pages} {260601} (\bibinfo {year} {2025})}\BibitemShut {NoStop}%
\bibitem [{\citenamefont {DiCarlo}\ \emph {et~al.}(2009)\citenamefont
  {DiCarlo}, \citenamefont {Chow}, \citenamefont {Gambetta}, \citenamefont
  {Bishop}, \citenamefont {Johnson}, \citenamefont {Schuster}, \citenamefont
  {Majer}, \citenamefont {Blais}, \citenamefont {Frunzio}, \citenamefont
  {Girvin},\ and\ \citenamefont {Schoelkopf}}]{dicarlo2009demonstration}%
  \BibitemOpen
  \bibfield  {author} {\bibinfo {author} {\bibfnamefont {L.}~\bibnamefont
  {DiCarlo}}, \bibinfo {author} {\bibfnamefont {J.~M.}\ \bibnamefont {Chow}},
  \bibinfo {author} {\bibfnamefont {J.~M.}\ \bibnamefont {Gambetta}}, \bibinfo
  {author} {\bibfnamefont {L.~S.}\ \bibnamefont {Bishop}}, \bibinfo {author}
  {\bibfnamefont {B.~R.}\ \bibnamefont {Johnson}}, \bibinfo {author}
  {\bibfnamefont {D.~I.}\ \bibnamefont {Schuster}}, \bibinfo {author}
  {\bibfnamefont {J.}~\bibnamefont {Majer}}, \bibinfo {author} {\bibfnamefont
  {A.}~\bibnamefont {Blais}}, \bibinfo {author} {\bibfnamefont
  {L.}~\bibnamefont {Frunzio}}, \bibinfo {author} {\bibfnamefont {S.~M.}\
  \bibnamefont {Girvin}},\ and\ \bibinfo {author} {\bibfnamefont {R.~J.}\
  \bibnamefont {Schoelkopf}},\ }\bibfield  {title} {\bibinfo {title}
  {Demonstration of two-qubit algorithms with a superconducting quantum
  processor},\ }\href {https://doi.org/10.1038/nature08121} {\bibfield
  {journal} {\bibinfo  {journal} {Nature}\ }\textbf {\bibinfo {volume} {460}},\
  \bibinfo {pages} {240} (\bibinfo {year} {2009})}\BibitemShut {NoStop}%
\bibitem [{\citenamefont {Chen}\ \emph {et~al.}(2014)\citenamefont {Chen},
  \citenamefont {Neill}, \citenamefont {Roushan}, \citenamefont {Leung},
  \citenamefont {Fang}, \citenamefont {Barends}, \citenamefont {Kelly},
  \citenamefont {Campbell}, \citenamefont {Chen}, \citenamefont {Chiaro},
  \citenamefont {Dunsworth}, \citenamefont {Jeffrey}, \citenamefont {Megrant},
  \citenamefont {Mutus}, \citenamefont {O'Malley}, \citenamefont {Quintana},
  \citenamefont {Sank}, \citenamefont {Vainsencher}, \citenamefont {Wenner},
  \citenamefont {White}, \citenamefont {Geller}, \citenamefont {Cleland},\ and\
  \citenamefont {Martinis}}]{PhysRevLett.113.220502}%
  \BibitemOpen
  \bibfield  {author} {\bibinfo {author} {\bibfnamefont {Y.}~\bibnamefont
  {Chen}}, \bibinfo {author} {\bibfnamefont {C.}~\bibnamefont {Neill}},
  \bibinfo {author} {\bibfnamefont {P.}~\bibnamefont {Roushan}}, \bibinfo
  {author} {\bibfnamefont {N.}~\bibnamefont {Leung}}, \bibinfo {author}
  {\bibfnamefont {M.}~\bibnamefont {Fang}}, \bibinfo {author} {\bibfnamefont
  {R.}~\bibnamefont {Barends}}, \bibinfo {author} {\bibfnamefont
  {J.}~\bibnamefont {Kelly}}, \bibinfo {author} {\bibfnamefont
  {B.}~\bibnamefont {Campbell}}, \bibinfo {author} {\bibfnamefont
  {Z.}~\bibnamefont {Chen}}, \bibinfo {author} {\bibfnamefont {B.}~\bibnamefont
  {Chiaro}}, \bibinfo {author} {\bibfnamefont {A.}~\bibnamefont {Dunsworth}},
  \bibinfo {author} {\bibfnamefont {E.}~\bibnamefont {Jeffrey}}, \bibinfo
  {author} {\bibfnamefont {A.}~\bibnamefont {Megrant}}, \bibinfo {author}
  {\bibfnamefont {J.~Y.}\ \bibnamefont {Mutus}}, \bibinfo {author}
  {\bibfnamefont {P.~J.~J.}\ \bibnamefont {O'Malley}}, \bibinfo {author}
  {\bibfnamefont {C.~M.}\ \bibnamefont {Quintana}}, \bibinfo {author}
  {\bibfnamefont {D.}~\bibnamefont {Sank}}, \bibinfo {author} {\bibfnamefont
  {A.}~\bibnamefont {Vainsencher}}, \bibinfo {author} {\bibfnamefont
  {J.}~\bibnamefont {Wenner}}, \bibinfo {author} {\bibfnamefont {T.~C.}\
  \bibnamefont {White}}, \bibinfo {author} {\bibfnamefont {M.~R.}\ \bibnamefont
  {Geller}}, \bibinfo {author} {\bibfnamefont {A.~N.}\ \bibnamefont
  {Cleland}},\ and\ \bibinfo {author} {\bibfnamefont {J.~M.}\ \bibnamefont
  {Martinis}},\ }\bibfield  {title} {\bibinfo {title} {Qubit architecture with
  high coherence and fast tunable coupling},\ }\href
  {https://doi.org/10.1103/PhysRevLett.113.220502} {\bibfield  {journal}
  {\bibinfo  {journal} {Phys. Rev. Lett.}\ }\textbf {\bibinfo {volume} {113}},\
  \bibinfo {pages} {220502} (\bibinfo {year} {2014})}\BibitemShut {NoStop}%
\bibitem [{\citenamefont {Yan}\ \emph {et~al.}(2018)\citenamefont {Yan},
  \citenamefont {Krantz}, \citenamefont {Sung}, \citenamefont {Kjaergaard},
  \citenamefont {Campbell}, \citenamefont {Orlando}, \citenamefont
  {Gustavsson},\ and\ \citenamefont {Oliver}}]{PhysRevApplied.10.054062}%
  \BibitemOpen
  \bibfield  {author} {\bibinfo {author} {\bibfnamefont {F.}~\bibnamefont
  {Yan}}, \bibinfo {author} {\bibfnamefont {P.}~\bibnamefont {Krantz}},
  \bibinfo {author} {\bibfnamefont {Y.}~\bibnamefont {Sung}}, \bibinfo {author}
  {\bibfnamefont {M.}~\bibnamefont {Kjaergaard}}, \bibinfo {author}
  {\bibfnamefont {D.~L.}\ \bibnamefont {Campbell}}, \bibinfo {author}
  {\bibfnamefont {T.~P.}\ \bibnamefont {Orlando}}, \bibinfo {author}
  {\bibfnamefont {S.}~\bibnamefont {Gustavsson}},\ and\ \bibinfo {author}
  {\bibfnamefont {W.~D.}\ \bibnamefont {Oliver}},\ }\bibfield  {title}
  {\bibinfo {title} {Tunable coupling scheme for implementing high-fidelity
  two-qubit gates},\ }\href {https://doi.org/10.1103/PhysRevApplied.10.054062}
  {\bibfield  {journal} {\bibinfo  {journal} {Phys. Rev. Appl.}\ }\textbf
  {\bibinfo {volume} {10}},\ \bibinfo {pages} {054062} (\bibinfo {year}
  {2018})}\BibitemShut {NoStop}%
\bibitem [{\citenamefont {Heunisch}\ \emph {et~al.}(2023)\citenamefont
  {Heunisch}, \citenamefont {Eichler},\ and\ \citenamefont
  {Hartmann}}]{PhysRevApplied.20.064037}%
  \BibitemOpen
  \bibfield  {author} {\bibinfo {author} {\bibfnamefont {L.}~\bibnamefont
  {Heunisch}}, \bibinfo {author} {\bibfnamefont {C.}~\bibnamefont {Eichler}},\
  and\ \bibinfo {author} {\bibfnamefont {M.~J.}\ \bibnamefont {Hartmann}},\
  }\bibfield  {title} {\bibinfo {title} {Tunable coupler to fully decouple and
  maximally localize superconducting qubits},\ }\href
  {https://doi.org/10.1103/PhysRevApplied.20.064037} {\bibfield  {journal}
  {\bibinfo  {journal} {Phys. Rev. Appl.}\ }\textbf {\bibinfo {volume} {20}},\
  \bibinfo {pages} {064037} (\bibinfo {year} {2023})}\BibitemShut {NoStop}%
\bibitem [{\citenamefont {Sung}\ \emph {et~al.}(2021)\citenamefont {Sung},
  \citenamefont {Ding}, \citenamefont {Braum\"uller}, \citenamefont
  {Veps\"al\"ainen}, \citenamefont {Kannan}, \citenamefont {Kjaergaard},
  \citenamefont {Greene}, \citenamefont {Samach}, \citenamefont {McNally},
  \citenamefont {Kim}, \citenamefont {Melville}, \citenamefont {Niedzielski},
  \citenamefont {Schwartz}, \citenamefont {Yoder}, \citenamefont {Orlando},
  \citenamefont {Gustavsson},\ and\ \citenamefont
  {Oliver}}]{PhysRevX.11.021058}%
  \BibitemOpen
  \bibfield  {author} {\bibinfo {author} {\bibfnamefont {Y.}~\bibnamefont
  {Sung}}, \bibinfo {author} {\bibfnamefont {L.}~\bibnamefont {Ding}}, \bibinfo
  {author} {\bibfnamefont {J.}~\bibnamefont {Braum\"uller}}, \bibinfo {author}
  {\bibfnamefont {A.}~\bibnamefont {Veps\"al\"ainen}}, \bibinfo {author}
  {\bibfnamefont {B.}~\bibnamefont {Kannan}}, \bibinfo {author} {\bibfnamefont
  {M.}~\bibnamefont {Kjaergaard}}, \bibinfo {author} {\bibfnamefont
  {A.}~\bibnamefont {Greene}}, \bibinfo {author} {\bibfnamefont {G.~O.}\
  \bibnamefont {Samach}}, \bibinfo {author} {\bibfnamefont {C.}~\bibnamefont
  {McNally}}, \bibinfo {author} {\bibfnamefont {D.}~\bibnamefont {Kim}},
  \bibinfo {author} {\bibfnamefont {A.}~\bibnamefont {Melville}}, \bibinfo
  {author} {\bibfnamefont {B.~M.}\ \bibnamefont {Niedzielski}}, \bibinfo
  {author} {\bibfnamefont {M.~E.}\ \bibnamefont {Schwartz}}, \bibinfo {author}
  {\bibfnamefont {J.~L.}\ \bibnamefont {Yoder}}, \bibinfo {author}
  {\bibfnamefont {T.~P.}\ \bibnamefont {Orlando}}, \bibinfo {author}
  {\bibfnamefont {S.}~\bibnamefont {Gustavsson}},\ and\ \bibinfo {author}
  {\bibfnamefont {W.~D.}\ \bibnamefont {Oliver}},\ }\bibfield  {title}
  {\bibinfo {title} {Realization of high-fidelity cz and $zz$-free iswap gates
  with a tunable coupler},\ }\href {https://doi.org/10.1103/PhysRevX.11.021058}
  {\bibfield  {journal} {\bibinfo  {journal} {Phys. Rev. X}\ }\textbf {\bibinfo
  {volume} {11}},\ \bibinfo {pages} {021058} (\bibinfo {year}
  {2021})}\BibitemShut {NoStop}%
\bibitem [{\citenamefont {Chang}\ \emph {et~al.}(2013)\citenamefont {Chang},
  \citenamefont {Vissers}, \citenamefont {C{\'o}rcoles}, \citenamefont
  {Sandberg}, \citenamefont {Gao}, \citenamefont {Abraham}, \citenamefont
  {Chow}, \citenamefont {Gambetta}, \citenamefont {Rothwell}, \citenamefont
  {Keefe}, \citenamefont {Steffen},\ and\ \citenamefont
  {Pappas}}]{Chang2013TiNCoherence}%
  \BibitemOpen
  \bibfield  {author} {\bibinfo {author} {\bibfnamefont {J.~B.}\ \bibnamefont
  {Chang}}, \bibinfo {author} {\bibfnamefont {M.~R.}\ \bibnamefont {Vissers}},
  \bibinfo {author} {\bibfnamefont {A.~D.}\ \bibnamefont {C{\'o}rcoles}},
  \bibinfo {author} {\bibfnamefont {M.}~\bibnamefont {Sandberg}}, \bibinfo
  {author} {\bibfnamefont {J.}~\bibnamefont {Gao}}, \bibinfo {author}
  {\bibfnamefont {D.~W.}\ \bibnamefont {Abraham}}, \bibinfo {author}
  {\bibfnamefont {J.~M.}\ \bibnamefont {Chow}}, \bibinfo {author}
  {\bibfnamefont {J.~M.}\ \bibnamefont {Gambetta}}, \bibinfo {author}
  {\bibfnamefont {M.~B.}\ \bibnamefont {Rothwell}}, \bibinfo {author}
  {\bibfnamefont {G.~A.}\ \bibnamefont {Keefe}}, \bibinfo {author}
  {\bibfnamefont {M.}~\bibnamefont {Steffen}},\ and\ \bibinfo {author}
  {\bibfnamefont {D.~P.}\ \bibnamefont {Pappas}},\ }\bibfield  {title}
  {\bibinfo {title} {Improved superconducting qubit coherence using titanium
  nitride},\ }\href {https://doi.org/10.1063/1.4813269} {\bibfield  {journal}
  {\bibinfo  {journal} {Applied Physics Letters}\ }\textbf {\bibinfo {volume}
  {103}},\ \bibinfo {pages} {012602} (\bibinfo {year} {2013})}\BibitemShut
  {NoStop}%
\bibitem [{\citenamefont {Place}\ \emph {et~al.}(2021)\citenamefont {Place},
  \citenamefont {Rodgers}, \citenamefont {Mundada}, \citenamefont {Smitham},
  \citenamefont {Fitzpatrick}, \citenamefont {Leng}, \citenamefont {Premkumar},
  \citenamefont {Bryon}, \citenamefont {Vrajitoarea}, \citenamefont {Sussman},
  \citenamefont {Cheng}, \citenamefont {Madhavan}, \citenamefont {Babla},
  \citenamefont {Le}, \citenamefont {Gang}, \citenamefont {Jäck},
  \citenamefont {Gyenis}, \citenamefont {Yao}, \citenamefont {Cava},
  \citenamefont {de~Leon},\ and\ \citenamefont
  {Houck}}]{Place2021TantalumTransmon}%
  \BibitemOpen
  \bibfield  {author} {\bibinfo {author} {\bibfnamefont {A.~P.~M.}\
  \bibnamefont {Place}}, \bibinfo {author} {\bibfnamefont {L.~V.~H.}\
  \bibnamefont {Rodgers}}, \bibinfo {author} {\bibfnamefont {P.}~\bibnamefont
  {Mundada}}, \bibinfo {author} {\bibfnamefont {B.~M.}\ \bibnamefont
  {Smitham}}, \bibinfo {author} {\bibfnamefont {M.}~\bibnamefont
  {Fitzpatrick}}, \bibinfo {author} {\bibfnamefont {Z.}~\bibnamefont {Leng}},
  \bibinfo {author} {\bibfnamefont {A.}~\bibnamefont {Premkumar}}, \bibinfo
  {author} {\bibfnamefont {J.}~\bibnamefont {Bryon}}, \bibinfo {author}
  {\bibfnamefont {A.}~\bibnamefont {Vrajitoarea}}, \bibinfo {author}
  {\bibfnamefont {S.}~\bibnamefont {Sussman}}, \bibinfo {author} {\bibfnamefont
  {G.}~\bibnamefont {Cheng}}, \bibinfo {author} {\bibfnamefont
  {T.}~\bibnamefont {Madhavan}}, \bibinfo {author} {\bibfnamefont {H.~K.}\
  \bibnamefont {Babla}}, \bibinfo {author} {\bibfnamefont {X.~H.}\ \bibnamefont
  {Le}}, \bibinfo {author} {\bibfnamefont {Y.}~\bibnamefont {Gang}}, \bibinfo
  {author} {\bibfnamefont {B.}~\bibnamefont {Jäck}}, \bibinfo {author}
  {\bibfnamefont {A.}~\bibnamefont {Gyenis}}, \bibinfo {author} {\bibfnamefont
  {N.}~\bibnamefont {Yao}}, \bibinfo {author} {\bibfnamefont {R.~J.}\
  \bibnamefont {Cava}}, \bibinfo {author} {\bibfnamefont {N.~P.}\ \bibnamefont
  {de~Leon}},\ and\ \bibinfo {author} {\bibfnamefont {A.~A.}\ \bibnamefont
  {Houck}},\ }\bibfield  {title} {\bibinfo {title} {New material platform for
  superconducting transmon qubits with coherence times exceeding 0.3
  milliseconds},\ }\href {https://doi.org/10.1038/s41467-021-22030-5}
  {\bibfield  {journal} {\bibinfo  {journal} {Nature Communications}\ }\textbf
  {\bibinfo {volume} {12}},\ \bibinfo {pages} {1779} (\bibinfo {year}
  {2021})}\BibitemShut {NoStop}%
\bibitem [{\citenamefont {Bland}\ \emph {et~al.}(2025)\citenamefont {Bland},
  \citenamefont {Bahrami}, \citenamefont {Martinez}, \citenamefont
  {Prestegaard}, \citenamefont {Smitham}, \citenamefont {Joshi}, \citenamefont
  {Hedrick}, \citenamefont {Kumar}, \citenamefont {Yang}, \citenamefont
  {Pakpour-Tabrizi}, \citenamefont {Jindal}, \citenamefont {Chang},
  \citenamefont {Cheng}, \citenamefont {Yao}, \citenamefont {Cava},
  \citenamefont {de~Leon},\ and\ \citenamefont
  {Houck}}]{Bland2025MillisecondTransmons}%
  \BibitemOpen
  \bibfield  {author} {\bibinfo {author} {\bibfnamefont {M.~P.}\ \bibnamefont
  {Bland}}, \bibinfo {author} {\bibfnamefont {F.}~\bibnamefont {Bahrami}},
  \bibinfo {author} {\bibfnamefont {J.~G.~C.}\ \bibnamefont {Martinez}},
  \bibinfo {author} {\bibfnamefont {P.~H.}\ \bibnamefont {Prestegaard}},
  \bibinfo {author} {\bibfnamefont {B.~M.}\ \bibnamefont {Smitham}}, \bibinfo
  {author} {\bibfnamefont {A.}~\bibnamefont {Joshi}}, \bibinfo {author}
  {\bibfnamefont {E.}~\bibnamefont {Hedrick}}, \bibinfo {author} {\bibfnamefont
  {S.}~\bibnamefont {Kumar}}, \bibinfo {author} {\bibfnamefont
  {A.}~\bibnamefont {Yang}}, \bibinfo {author} {\bibfnamefont {A.~C.}\
  \bibnamefont {Pakpour-Tabrizi}}, \bibinfo {author} {\bibfnamefont
  {A.}~\bibnamefont {Jindal}}, \bibinfo {author} {\bibfnamefont {R.~D.}\
  \bibnamefont {Chang}}, \bibinfo {author} {\bibfnamefont {G.}~\bibnamefont
  {Cheng}}, \bibinfo {author} {\bibfnamefont {N.}~\bibnamefont {Yao}}, \bibinfo
  {author} {\bibfnamefont {R.~J.}\ \bibnamefont {Cava}}, \bibinfo {author}
  {\bibfnamefont {N.~P.}\ \bibnamefont {de~Leon}},\ and\ \bibinfo {author}
  {\bibfnamefont {A.~A.}\ \bibnamefont {Houck}},\ }\bibfield  {title} {\bibinfo
  {title} {Millisecond lifetimes and coherence times in 2d transmon qubits},\
  }\href {https://doi.org/10.1038/s41586-025-09687-4} {\bibfield  {journal}
  {\bibinfo  {journal} {Nature}\ }\textbf {\bibinfo {volume} {647}},\ \bibinfo
  {pages} {343} (\bibinfo {year} {2025})}\BibitemShut {NoStop}%
\bibitem [{\citenamefont {Smirnov}\ \emph {et~al.}(2025)\citenamefont {Smirnov}
  \emph {et~al.}}]{Smirnov:2025cff}%
  \BibitemOpen
  \bibfield  {author} {\bibinfo {author} {\bibfnamefont {N.~S.}\ \bibnamefont
  {Smirnov}} \emph {et~al.},\ }\href@noop {} {\bibinfo {title} {{High-fidelity
  two-qubit gates with transmon qubits using bipolar flux pulses and tunable
  couplers}}} (\bibinfo {year} {2025}),\ \Eprint
  {https://arxiv.org/abs/2509.04965} {arXiv:2509.04965 [quant-ph]} \BibitemShut
  {NoStop}%
\bibitem [{\citenamefont {Chen}\ \emph {et~al.}(2026)\citenamefont {Chen},
  \citenamefont {Wu}, \citenamefont {Strong},\ and\ \citenamefont
  {Poletto}}]{chen2026unlocking}%
  \BibitemOpen
  \bibfield  {author} {\bibinfo {author} {\bibfnamefont {A.~Q.}\ \bibnamefont
  {Chen}}, \bibinfo {author} {\bibfnamefont {X.}~\bibnamefont {Wu}}, \bibinfo
  {author} {\bibfnamefont {S.}~\bibnamefont {Strong}},\ and\ \bibinfo {author}
  {\bibfnamefont {S.}~\bibnamefont {Poletto}},\ }\href@noop {} {\bibinfo
  {title} {Unlocking a fast adiabatic cz gate and exact residual zz
  cancellation between fixed-frequency transmons using a floating tunable
  coupler}} (\bibinfo {year} {2026}),\ \Eprint
  {https://arxiv.org/abs/2604.05048} {arXiv:2604.05048 [quant-ph]} \BibitemShut
  {NoStop}%
\bibitem [{\citenamefont {Li}\ \emph {et~al.}(2025)\citenamefont {Li},
  \citenamefont {Zhang}, \citenamefont {Chen}, \citenamefont {Huang},
  \citenamefont {Liu}, \citenamefont {Xiao}, \citenamefont {Deng},
  \citenamefont {Liang}, \citenamefont {Chen}, \citenamefont {Liu},
  \citenamefont {Li}, \citenamefont {Bao}, \citenamefont {Zhao}, \citenamefont
  {Xu}, \citenamefont {Li}, \citenamefont {He}, \citenamefont {Liu},
  \citenamefont {Yu}, \citenamefont {Zhou}, \citenamefont {Liu}, \citenamefont
  {Song}, \citenamefont {Zheng}, \citenamefont {Xiang}, \citenamefont {Shi},
  \citenamefont {Xu},\ and\ \citenamefont {Fan}}]{PhysRevApplied.23.024059}%
  \BibitemOpen
  \bibfield  {author} {\bibinfo {author} {\bibfnamefont {T.-M.}\ \bibnamefont
  {Li}}, \bibinfo {author} {\bibfnamefont {J.-C.}\ \bibnamefont {Zhang}},
  \bibinfo {author} {\bibfnamefont {B.-J.}\ \bibnamefont {Chen}}, \bibinfo
  {author} {\bibfnamefont {K.}~\bibnamefont {Huang}}, \bibinfo {author}
  {\bibfnamefont {H.-T.}\ \bibnamefont {Liu}}, \bibinfo {author} {\bibfnamefont
  {Y.-X.}\ \bibnamefont {Xiao}}, \bibinfo {author} {\bibfnamefont {C.-L.}\
  \bibnamefont {Deng}}, \bibinfo {author} {\bibfnamefont {G.-H.}\ \bibnamefont
  {Liang}}, \bibinfo {author} {\bibfnamefont {C.-T.}\ \bibnamefont {Chen}},
  \bibinfo {author} {\bibfnamefont {Y.}~\bibnamefont {Liu}}, \bibinfo {author}
  {\bibfnamefont {H.}~\bibnamefont {Li}}, \bibinfo {author} {\bibfnamefont
  {Z.-T.}\ \bibnamefont {Bao}}, \bibinfo {author} {\bibfnamefont
  {K.}~\bibnamefont {Zhao}}, \bibinfo {author} {\bibfnamefont {Y.}~\bibnamefont
  {Xu}}, \bibinfo {author} {\bibfnamefont {L.}~\bibnamefont {Li}}, \bibinfo
  {author} {\bibfnamefont {Y.}~\bibnamefont {He}}, \bibinfo {author}
  {\bibfnamefont {Z.-H.}\ \bibnamefont {Liu}}, \bibinfo {author} {\bibfnamefont
  {Y.-H.}\ \bibnamefont {Yu}}, \bibinfo {author} {\bibfnamefont {S.-Y.}\
  \bibnamefont {Zhou}}, \bibinfo {author} {\bibfnamefont {Y.-J.}\ \bibnamefont
  {Liu}}, \bibinfo {author} {\bibfnamefont {X.}~\bibnamefont {Song}}, \bibinfo
  {author} {\bibfnamefont {D.}~\bibnamefont {Zheng}}, \bibinfo {author}
  {\bibfnamefont {Z.}~\bibnamefont {Xiang}}, \bibinfo {author} {\bibfnamefont
  {Y.-H.}\ \bibnamefont {Shi}}, \bibinfo {author} {\bibfnamefont
  {K.}~\bibnamefont {Xu}},\ and\ \bibinfo {author} {\bibfnamefont
  {H.}~\bibnamefont {Fan}},\ }\bibfield  {title} {\bibinfo {title}
  {High-precision pulse calibration of tunable couplers for high-fidelity
  two-qubit gates in superconducting quantum processors},\ }\href
  {https://doi.org/10.1103/PhysRevApplied.23.024059} {\bibfield  {journal}
  {\bibinfo  {journal} {Phys. Rev. Appl.}\ }\textbf {\bibinfo {volume} {23}},\
  \bibinfo {pages} {024059} (\bibinfo {year} {2025})}\BibitemShut {NoStop}%
\bibitem [{\citenamefont {Ding}\ \emph {et~al.}(2025)\citenamefont {Ding},
  \citenamefont {Oppenheim}, \citenamefont {Boufounos}, \citenamefont
  {Gustavsson}, \citenamefont {Grover}, \citenamefont {Baran},\ and\
  \citenamefont {Oliver}}]{Ding2025PulseDesign}%
  \BibitemOpen
  \bibfield  {author} {\bibinfo {author} {\bibfnamefont {Q.}~\bibnamefont
  {Ding}}, \bibinfo {author} {\bibfnamefont {A.~V.}\ \bibnamefont {Oppenheim}},
  \bibinfo {author} {\bibfnamefont {P.~T.}\ \bibnamefont {Boufounos}}, \bibinfo
  {author} {\bibfnamefont {S.}~\bibnamefont {Gustavsson}}, \bibinfo {author}
  {\bibfnamefont {J.~A.}\ \bibnamefont {Grover}}, \bibinfo {author}
  {\bibfnamefont {T.~A.}\ \bibnamefont {Baran}},\ and\ \bibinfo {author}
  {\bibfnamefont {W.~D.}\ \bibnamefont {Oliver}},\ }\bibfield  {title}
  {\bibinfo {title} {Pulse design of baseband flux control for adiabatic
  controlled-phase gates in superconducting circuits},\ }\href
  {https://doi.org/10.1103/PhysRevApplied.23.064013} {\bibfield  {journal}
  {\bibinfo  {journal} {Physical Review Applied}\ }\textbf {\bibinfo {volume}
  {23}},\ \bibinfo {pages} {064013} (\bibinfo {year} {2025})}\BibitemShut
  {NoStop}%
\bibitem [{\citenamefont {Motzoi}\ \emph {et~al.}(2009)\citenamefont {Motzoi},
  \citenamefont {Gambetta}, \citenamefont {Rebentrost},\ and\ \citenamefont
  {Wilhelm}}]{PhysRevLett.103.110501}%
  \BibitemOpen
  \bibfield  {author} {\bibinfo {author} {\bibfnamefont {F.}~\bibnamefont
  {Motzoi}}, \bibinfo {author} {\bibfnamefont {J.~M.}\ \bibnamefont
  {Gambetta}}, \bibinfo {author} {\bibfnamefont {P.}~\bibnamefont
  {Rebentrost}},\ and\ \bibinfo {author} {\bibfnamefont {F.~K.}\ \bibnamefont
  {Wilhelm}},\ }\bibfield  {title} {\bibinfo {title} {Simple pulses for
  elimination of leakage in weakly nonlinear qubits},\ }\href
  {https://doi.org/10.1103/PhysRevLett.103.110501} {\bibfield  {journal}
  {\bibinfo  {journal} {Phys. Rev. Lett.}\ }\textbf {\bibinfo {volume} {103}},\
  \bibinfo {pages} {110501} (\bibinfo {year} {2009})}\BibitemShut {NoStop}%
\bibitem [{\citenamefont {Motzoi}\ and\ \citenamefont
  {Wilhelm}(2013)}]{PhysRevA.88.062318}%
  \BibitemOpen
  \bibfield  {author} {\bibinfo {author} {\bibfnamefont {F.}~\bibnamefont
  {Motzoi}}\ and\ \bibinfo {author} {\bibfnamefont {F.~K.}\ \bibnamefont
  {Wilhelm}},\ }\bibfield  {title} {\bibinfo {title} {Improving frequency
  selection of driven pulses using derivative-based transition suppression},\
  }\href {https://doi.org/10.1103/PhysRevA.88.062318} {\bibfield  {journal}
  {\bibinfo  {journal} {Phys. Rev. A}\ }\textbf {\bibinfo {volume} {88}},\
  \bibinfo {pages} {062318} (\bibinfo {year} {2013})}\BibitemShut {NoStop}%
\bibitem [{\citenamefont {Jesus}\ \emph {et~al.}(2025)\citenamefont {Jesus},
  \citenamefont {Li}, \citenamefont {Gao}, \citenamefont {Barends},
  \citenamefont {Cárdenas-López},\ and\ \citenamefont
  {Motzoi}}]{Jesus2025AnalyticalXGates}%
  \BibitemOpen
  \bibfield  {author} {\bibinfo {author} {\bibfnamefont {J.~D. D.~C.}\
  \bibnamefont {Jesus}}, \bibinfo {author} {\bibfnamefont {B.}~\bibnamefont
  {Li}}, \bibinfo {author} {\bibfnamefont {Y.}~\bibnamefont {Gao}}, \bibinfo
  {author} {\bibfnamefont {R.}~\bibnamefont {Barends}}, \bibinfo {author}
  {\bibfnamefont {F.~A.}\ \bibnamefont {Cárdenas-López}},\ and\ \bibinfo
  {author} {\bibfnamefont {F.}~\bibnamefont {Motzoi}},\ }\href@noop {}
  {\bibinfo {title} {Analytical blueprint for 99.999\% fidelity x-gates on
  present superconducting hardware under strong driving}} (\bibinfo {year}
  {2025}),\ \Eprint {https://arxiv.org/abs/2512.19919} {arXiv:2512.19919
  [quant-ph]} \BibitemShut {NoStop}%
\bibitem [{\citenamefont {Hyypp\"a}\ \emph {et~al.}(2024)\citenamefont
  {Hyypp\"a}, \citenamefont {Veps\"al\"ainen}, \citenamefont
  {Papi\ifmmode~\check{c}\else \v{c}\fi{}}, \citenamefont {Chan}, \citenamefont
  {Inel}, \citenamefont {Landra}, \citenamefont {Liu}, \citenamefont {Luus},
  \citenamefont {Marxer}, \citenamefont {Ockeloen-Korppi}, \citenamefont
  {Orbell}, \citenamefont {Tarasinski},\ and\ \citenamefont
  {Heinsoo}}]{PRXQuantum.5.030353}%
  \BibitemOpen
  \bibfield  {author} {\bibinfo {author} {\bibfnamefont {E.}~\bibnamefont
  {Hyypp\"a}}, \bibinfo {author} {\bibfnamefont {A.}~\bibnamefont
  {Veps\"al\"ainen}}, \bibinfo {author} {\bibfnamefont {M.}~\bibnamefont
  {Papi\ifmmode~\check{c}\else \v{c}\fi{}}}, \bibinfo {author} {\bibfnamefont
  {C.~F.}\ \bibnamefont {Chan}}, \bibinfo {author} {\bibfnamefont
  {S.}~\bibnamefont {Inel}}, \bibinfo {author} {\bibfnamefont {A.}~\bibnamefont
  {Landra}}, \bibinfo {author} {\bibfnamefont {W.}~\bibnamefont {Liu}},
  \bibinfo {author} {\bibfnamefont {J.}~\bibnamefont {Luus}}, \bibinfo {author}
  {\bibfnamefont {F.}~\bibnamefont {Marxer}}, \bibinfo {author} {\bibfnamefont
  {C.}~\bibnamefont {Ockeloen-Korppi}}, \bibinfo {author} {\bibfnamefont
  {S.}~\bibnamefont {Orbell}}, \bibinfo {author} {\bibfnamefont
  {B.}~\bibnamefont {Tarasinski}},\ and\ \bibinfo {author} {\bibfnamefont
  {J.}~\bibnamefont {Heinsoo}},\ }\bibfield  {title} {\bibinfo {title}
  {Reducing leakage of single-qubit gates for superconducting quantum
  processors using analytical control pulse envelopes},\ }\href
  {https://doi.org/10.1103/PRXQuantum.5.030353} {\bibfield  {journal} {\bibinfo
   {journal} {PRX Quantum}\ }\textbf {\bibinfo {volume} {5}},\ \bibinfo {pages}
  {030353} (\bibinfo {year} {2024})}\BibitemShut {NoStop}%
\bibitem [{\citenamefont {Li}\ \emph {et~al.}(2024)\citenamefont {Li},
  \citenamefont {Calarco},\ and\ \citenamefont
  {Motzoi}}]{Li2024CRErrorSuppression}%
  \BibitemOpen
  \bibfield  {author} {\bibinfo {author} {\bibfnamefont {B.}~\bibnamefont
  {Li}}, \bibinfo {author} {\bibfnamefont {T.}~\bibnamefont {Calarco}},\ and\
  \bibinfo {author} {\bibfnamefont {F.}~\bibnamefont {Motzoi}},\ }\bibfield
  {title} {\bibinfo {title} {Experimental error suppression in cross-resonance
  gates via multi-derivative pulse shaping},\ }\href
  {https://doi.org/10.1038/s41534-024-00863-4} {\bibfield  {journal} {\bibinfo
  {journal} {npj Quantum Information}\ }\textbf {\bibinfo {volume} {10}},\
  \bibinfo {pages} {66} (\bibinfo {year} {2024})}\BibitemShut {NoStop}%
\bibitem [{\citenamefont {Gao}\ \emph {et~al.}(2025)\citenamefont {Gao},
  \citenamefont {Galicia}, \citenamefont {Da~Costa~Jesus}, \citenamefont {Liu},
  \citenamefont {Haddad}, \citenamefont {Volkov}, \citenamefont
  {Guimar{\~a}es}, \citenamefont {Bhardwaj}, \citenamefont {Jerger},
  \citenamefont {Neis}, \citenamefont {Li}, \citenamefont
  {C{\'a}rdenas-L{\'o}pez}, \citenamefont {Motzoi}, \citenamefont {Bushev},\
  and\ \citenamefont {Barends}}]{Gao2025UltrafastSingleQubit}%
  \BibitemOpen
  \bibfield  {author} {\bibinfo {author} {\bibfnamefont {Y.}~\bibnamefont
  {Gao}}, \bibinfo {author} {\bibfnamefont {A.}~\bibnamefont {Galicia}},
  \bibinfo {author} {\bibfnamefont {J.~D.}\ \bibnamefont {Da~Costa~Jesus}},
  \bibinfo {author} {\bibfnamefont {Y.}~\bibnamefont {Liu}}, \bibinfo {author}
  {\bibfnamefont {Y.}~\bibnamefont {Haddad}}, \bibinfo {author} {\bibfnamefont
  {D.~A.}\ \bibnamefont {Volkov}}, \bibinfo {author} {\bibfnamefont {J.~R.}\
  \bibnamefont {Guimar{\~a}es}}, \bibinfo {author} {\bibfnamefont
  {H.}~\bibnamefont {Bhardwaj}}, \bibinfo {author} {\bibfnamefont
  {M.}~\bibnamefont {Jerger}}, \bibinfo {author} {\bibfnamefont
  {M.}~\bibnamefont {Neis}}, \bibinfo {author} {\bibfnamefont {B.}~\bibnamefont
  {Li}}, \bibinfo {author} {\bibfnamefont {F.~A.}\ \bibnamefont
  {C{\'a}rdenas-L{\'o}pez}}, \bibinfo {author} {\bibfnamefont {F.}~\bibnamefont
  {Motzoi}}, \bibinfo {author} {\bibfnamefont {P.~A.}\ \bibnamefont {Bushev}},\
  and\ \bibinfo {author} {\bibfnamefont {R.}~\bibnamefont {Barends}},\
  }\href@noop {} {\bibinfo {title} {Ultrafast single qubit gates through
  multi-photon transition removal}} (\bibinfo {year} {2025}),\ \Eprint
  {https://arxiv.org/abs/2511.22365} {arXiv:2511.22365 [quant-ph]} \BibitemShut
  {NoStop}%
\bibitem [{\citenamefont {Koch}\ \emph {et~al.}(2007)\citenamefont {Koch},
  \citenamefont {Yu}, \citenamefont {Gambetta}, \citenamefont {Houck},
  \citenamefont {Schuster}, \citenamefont {Majer}, \citenamefont {Blais},
  \citenamefont {Devoret}, \citenamefont {Girvin},\ and\ \citenamefont
  {Schoelkopf}}]{PhysRevA.76.042319}%
  \BibitemOpen
  \bibfield  {author} {\bibinfo {author} {\bibfnamefont {J.}~\bibnamefont
  {Koch}}, \bibinfo {author} {\bibfnamefont {T.~M.}\ \bibnamefont {Yu}},
  \bibinfo {author} {\bibfnamefont {J.}~\bibnamefont {Gambetta}}, \bibinfo
  {author} {\bibfnamefont {A.~A.}\ \bibnamefont {Houck}}, \bibinfo {author}
  {\bibfnamefont {D.~I.}\ \bibnamefont {Schuster}}, \bibinfo {author}
  {\bibfnamefont {J.}~\bibnamefont {Majer}}, \bibinfo {author} {\bibfnamefont
  {A.}~\bibnamefont {Blais}}, \bibinfo {author} {\bibfnamefont {M.~H.}\
  \bibnamefont {Devoret}}, \bibinfo {author} {\bibfnamefont {S.~M.}\
  \bibnamefont {Girvin}},\ and\ \bibinfo {author} {\bibfnamefont {R.~J.}\
  \bibnamefont {Schoelkopf}},\ }\bibfield  {title} {\bibinfo {title}
  {Charge-insensitive qubit design derived from the cooper pair box},\ }\href
  {https://doi.org/10.1103/PhysRevA.76.042319} {\bibfield  {journal} {\bibinfo
  {journal} {Phys. Rev. A}\ }\textbf {\bibinfo {volume} {76}},\ \bibinfo
  {pages} {042319} (\bibinfo {year} {2007})}\BibitemShut {NoStop}%
\bibitem [{\citenamefont {Goerz}\ \emph {et~al.}(2017)\citenamefont {Goerz},
  \citenamefont {Motzoi}, \citenamefont {Whaley},\ and\ \citenamefont
  {Koch}}]{goerz_charting_2017}%
  \BibitemOpen
  \bibfield  {author} {\bibinfo {author} {\bibfnamefont {M.~H.}\ \bibnamefont
  {Goerz}}, \bibinfo {author} {\bibfnamefont {F.}~\bibnamefont {Motzoi}},
  \bibinfo {author} {\bibfnamefont {K.~B.}\ \bibnamefont {Whaley}},\ and\
  \bibinfo {author} {\bibfnamefont {C.~P.}\ \bibnamefont {Koch}},\ }\bibfield
  {title} {\bibinfo {title} {Charting the circuit {QED} design landscape using
  optimal control theory},\ }\href {https://doi.org/10.1038/s41534-017-0036-0}
  {\bibfield  {journal} {\bibinfo  {journal} {npj Quantum Information}\
  }\textbf {\bibinfo {volume} {3}},\ \bibinfo {pages} {37} (\bibinfo {year}
  {2017})}\BibitemShut {NoStop}%
\bibitem [{\citenamefont {Schrieffer}\ and\ \citenamefont
  {Wolff}(1966)}]{PhysRev.149.491}%
  \BibitemOpen
  \bibfield  {author} {\bibinfo {author} {\bibfnamefont {J.~R.}\ \bibnamefont
  {Schrieffer}}\ and\ \bibinfo {author} {\bibfnamefont {P.~A.}\ \bibnamefont
  {Wolff}},\ }\bibfield  {title} {\bibinfo {title} {Relation between the
  anderson and kondo hamiltonians},\ }\href
  {https://doi.org/10.1103/PhysRev.149.491} {\bibfield  {journal} {\bibinfo
  {journal} {Phys. Rev.}\ }\textbf {\bibinfo {volume} {149}},\ \bibinfo {pages}
  {491} (\bibinfo {year} {1966})}\BibitemShut {NoStop}%
\bibitem [{\citenamefont {Wu}\ and\ \citenamefont
  {Yang}(2007)}]{PhysRevLett.98.013601}%
  \BibitemOpen
  \bibfield  {author} {\bibinfo {author} {\bibfnamefont {Y.}~\bibnamefont
  {Wu}}\ and\ \bibinfo {author} {\bibfnamefont {X.}~\bibnamefont {Yang}},\
  }\bibfield  {title} {\bibinfo {title} {Strong-coupling theory of periodically
  driven two-level systems},\ }\href
  {https://doi.org/10.1103/PhysRevLett.98.013601} {\bibfield  {journal}
  {\bibinfo  {journal} {Phys. Rev. Lett.}\ }\textbf {\bibinfo {volume} {98}},\
  \bibinfo {pages} {013601} (\bibinfo {year} {2007})}\BibitemShut {NoStop}%
\bibitem [{\citenamefont {Watts}\ \emph {et~al.}(2015)\citenamefont {Watts},
  \citenamefont {Vala}, \citenamefont {M\"uller}, \citenamefont {Calarco},
  \citenamefont {Whaley}, \citenamefont {Reich}, \citenamefont {Goerz},\ and\
  \citenamefont {Koch}}]{PhysRevA.91.062306}%
  \BibitemOpen
  \bibfield  {author} {\bibinfo {author} {\bibfnamefont {P.}~\bibnamefont
  {Watts}}, \bibinfo {author} {\bibfnamefont {J.~c.~v.}\ \bibnamefont {Vala}},
  \bibinfo {author} {\bibfnamefont {M.~M.}\ \bibnamefont {M\"uller}}, \bibinfo
  {author} {\bibfnamefont {T.}~\bibnamefont {Calarco}}, \bibinfo {author}
  {\bibfnamefont {K.~B.}\ \bibnamefont {Whaley}}, \bibinfo {author}
  {\bibfnamefont {D.~M.}\ \bibnamefont {Reich}}, \bibinfo {author}
  {\bibfnamefont {M.~H.}\ \bibnamefont {Goerz}},\ and\ \bibinfo {author}
  {\bibfnamefont {C.~P.}\ \bibnamefont {Koch}},\ }\bibfield  {title} {\bibinfo
  {title} {Optimizing for an arbitrary perfect entangler. i. functionals},\
  }\href {https://doi.org/10.1103/PhysRevA.91.062306} {\bibfield  {journal}
  {\bibinfo  {journal} {Phys. Rev. A}\ }\textbf {\bibinfo {volume} {91}},\
  \bibinfo {pages} {062306} (\bibinfo {year} {2015})}\BibitemShut {NoStop}%
\bibitem [{\citenamefont {Groszkowski}\ \emph {et~al.}(2011)\citenamefont
  {Groszkowski}, \citenamefont {Fowler}, \citenamefont {Motzoi},\ and\
  \citenamefont {Wilhelm}}]{groszkowski2011tunable}%
  \BibitemOpen
  \bibfield  {author} {\bibinfo {author} {\bibfnamefont {P.}~\bibnamefont
  {Groszkowski}}, \bibinfo {author} {\bibfnamefont {A.~G.}\ \bibnamefont
  {Fowler}}, \bibinfo {author} {\bibfnamefont {F.}~\bibnamefont {Motzoi}},\
  and\ \bibinfo {author} {\bibfnamefont {F.~K.}\ \bibnamefont {Wilhelm}},\
  }\bibfield  {title} {\bibinfo {title} {Tunable coupling between three qubits
  as a building block for a superconducting quantum computer},\ }\href
  {https://doi.org/10.1103/PhysRevB.84.144516} {\bibfield  {journal} {\bibinfo
  {journal} {Phys. Rev. B}\ }\textbf {\bibinfo {volume} {84}},\ \bibinfo
  {pages} {144516} (\bibinfo {year} {2011})}\BibitemShut {NoStop}%
\bibitem [{\citenamefont {Theis}\ \emph {et~al.}(2016)\citenamefont {Theis},
  \citenamefont {Motzoi},\ and\ \citenamefont {Wilhelm}}]{theis2016}%
  \BibitemOpen
  \bibfield  {author} {\bibinfo {author} {\bibfnamefont {L.~S.}\ \bibnamefont
  {Theis}}, \bibinfo {author} {\bibfnamefont {F.}~\bibnamefont {Motzoi}},\ and\
  \bibinfo {author} {\bibfnamefont {F.~K.}\ \bibnamefont {Wilhelm}},\
  }\bibfield  {title} {\bibinfo {title} {Simultaneous gates in
  frequency-crowded multilevel systems using fast, robust, analytic control
  shapes},\ }\href {https://doi.org/10.1103/PhysRevA.93.012324} {\bibfield
  {journal} {\bibinfo  {journal} {Phys. Rev. A}\ }\textbf {\bibinfo {volume}
  {93}},\ \bibinfo {pages} {012324} (\bibinfo {year} {2016})}\BibitemShut
  {NoStop}%
\bibitem [{\citenamefont {Wang}\ \emph {et~al.}(2025)\citenamefont {Wang},
  \citenamefont {Feng}, \citenamefont {Zhang}, \citenamefont {Ding},
  \citenamefont {Li}, \citenamefont {Motzoi}, \citenamefont {Gao},
  \citenamefont {Xu}, \citenamefont {Yang}, \citenamefont {Nuerbolati},
  \citenamefont {Yu}, \citenamefont {Sun},\ and\ \citenamefont
  {Yan}}]{h4xf-vq2l}%
  \BibitemOpen
  \bibfield  {author} {\bibinfo {author} {\bibfnamefont {R.}~\bibnamefont
  {Wang}}, \bibinfo {author} {\bibfnamefont {Y.}~\bibnamefont {Feng}}, \bibinfo
  {author} {\bibfnamefont {Y.}~\bibnamefont {Zhang}}, \bibinfo {author}
  {\bibfnamefont {J.}~\bibnamefont {Ding}}, \bibinfo {author} {\bibfnamefont
  {B.}~\bibnamefont {Li}}, \bibinfo {author} {\bibfnamefont {F.}~\bibnamefont
  {Motzoi}}, \bibinfo {author} {\bibfnamefont {Y.}~\bibnamefont {Gao}},
  \bibinfo {author} {\bibfnamefont {H.}~\bibnamefont {Xu}}, \bibinfo {author}
  {\bibfnamefont {Z.}~\bibnamefont {Yang}}, \bibinfo {author} {\bibfnamefont
  {W.}~\bibnamefont {Nuerbolati}}, \bibinfo {author} {\bibfnamefont
  {H.}~\bibnamefont {Yu}}, \bibinfo {author} {\bibfnamefont {W.}~\bibnamefont
  {Sun}},\ and\ \bibinfo {author} {\bibfnamefont {F.}~\bibnamefont {Yan}},\
  }\bibfield  {title} {\bibinfo {title} {Suppressing spurious transitions using
  spectrally balanced pulse},\ }\href {https://doi.org/10.1103/h4xf-vq2l}
  {\bibfield  {journal} {\bibinfo  {journal} {Phys. Rev. Lett.}\ }\textbf
  {\bibinfo {volume} {135}},\ \bibinfo {pages} {160804} (\bibinfo {year}
  {2025})}\BibitemShut {NoStop}%
\bibitem [{\citenamefont {Hellings}\ \emph {et~al.}(2025)\citenamefont
  {Hellings}, \citenamefont {Lacroix}, \citenamefont {Remm}, \citenamefont
  {Boell}, \citenamefont {Herrmann}, \citenamefont {Laz\ifmmode~\u{a}\else
  \u{a}\fi{}r}, \citenamefont {Krinner}, \citenamefont {Swiadek}, \citenamefont
  {Andersen}, \citenamefont {Eichler},\ and\ \citenamefont
  {Wallraff}}]{1qhb-r4fb}%
  \BibitemOpen
  \bibfield  {author} {\bibinfo {author} {\bibfnamefont {C.}~\bibnamefont
  {Hellings}}, \bibinfo {author} {\bibfnamefont {N.}~\bibnamefont {Lacroix}},
  \bibinfo {author} {\bibfnamefont {A.}~\bibnamefont {Remm}}, \bibinfo {author}
  {\bibfnamefont {R.}~\bibnamefont {Boell}}, \bibinfo {author} {\bibfnamefont
  {J.}~\bibnamefont {Herrmann}}, \bibinfo {author} {\bibfnamefont
  {S.}~\bibnamefont {Laz\ifmmode~\u{a}\else \u{a}\fi{}r}}, \bibinfo {author}
  {\bibfnamefont {S.}~\bibnamefont {Krinner}}, \bibinfo {author} {\bibfnamefont
  {F.~m.~c.}\ \bibnamefont {Swiadek}}, \bibinfo {author} {\bibfnamefont
  {C.~K.}\ \bibnamefont {Andersen}}, \bibinfo {author} {\bibfnamefont
  {C.}~\bibnamefont {Eichler}},\ and\ \bibinfo {author} {\bibfnamefont
  {A.}~\bibnamefont {Wallraff}},\ }\bibfield  {title} {\bibinfo {title}
  {Calibrating magnetic flux control in superconducting circuits by
  compensating distortions on timescales from nanoseconds up to tens of
  microseconds},\ }\href {https://doi.org/10.1103/1qhb-r4fb} {\bibfield
  {journal} {\bibinfo  {journal} {Phys. Rev. Res.}\ }\textbf {\bibinfo {volume}
  {7}},\ \bibinfo {pages} {043142} (\bibinfo {year} {2025})}\BibitemShut
  {NoStop}%
\bibitem [{\citenamefont {Rol}\ \emph {et~al.}(2020)\citenamefont {Rol},
  \citenamefont {Ciorciaro}, \citenamefont {Malinowski}, \citenamefont
  {Tarasinski}, \citenamefont {Sagastizabal}, \citenamefont {Bultink},
  \citenamefont {Salathe}, \citenamefont {Haandbaek}, \citenamefont {Sedivy},\
  and\ \citenamefont {DiCarlo}}]{rol2020cryoscope}%
  \BibitemOpen
  \bibfield  {author} {\bibinfo {author} {\bibfnamefont {M.~A.}\ \bibnamefont
  {Rol}}, \bibinfo {author} {\bibfnamefont {L.}~\bibnamefont {Ciorciaro}},
  \bibinfo {author} {\bibfnamefont {F.~K.}\ \bibnamefont {Malinowski}},
  \bibinfo {author} {\bibfnamefont {B.~M.}\ \bibnamefont {Tarasinski}},
  \bibinfo {author} {\bibfnamefont {R.~E.}\ \bibnamefont {Sagastizabal}},
  \bibinfo {author} {\bibfnamefont {C.~C.}\ \bibnamefont {Bultink}}, \bibinfo
  {author} {\bibfnamefont {Y.}~\bibnamefont {Salathe}}, \bibinfo {author}
  {\bibfnamefont {N.}~\bibnamefont {Haandbaek}}, \bibinfo {author}
  {\bibfnamefont {J.}~\bibnamefont {Sedivy}},\ and\ \bibinfo {author}
  {\bibfnamefont {L.}~\bibnamefont {DiCarlo}},\ }\bibfield  {title} {\bibinfo
  {title} {Time-domain characterization and correction of on-chip distortion of
  control pulses in a quantum processor},\ }\href
  {https://doi.org/10.1063/1.5133894} {\bibfield  {journal} {\bibinfo
  {journal} {Applied Physics Letters}\ }\textbf {\bibinfo {volume} {116}},\
  \bibinfo {pages} {054001} (\bibinfo {year} {2020})}\BibitemShut {NoStop}%
\bibitem [{\citenamefont {Barends}\ \emph {et~al.}(2014)\citenamefont
  {Barends}, \citenamefont {Kelly}, \citenamefont {Megrant}, \citenamefont
  {Veitia}, \citenamefont {Sank}, \citenamefont {Jeffrey}, \citenamefont
  {White}, \citenamefont {Mutus}, \citenamefont {Fowler}, \citenamefont
  {Campbell}, \citenamefont {Chen}, \citenamefont {Chen}, \citenamefont
  {Chiaro}, \citenamefont {Dunsworth}, \citenamefont {Neill}, \citenamefont
  {O'Malley}, \citenamefont {Roushan}, \citenamefont {Vainsencher},
  \citenamefont {Wenner}, \citenamefont {Korotkov}, \citenamefont {Cleland},\
  and\ \citenamefont {Martinis}}]{barends2014surfacecode}%
  \BibitemOpen
  \bibfield  {author} {\bibinfo {author} {\bibfnamefont {R.}~\bibnamefont
  {Barends}}, \bibinfo {author} {\bibfnamefont {J.}~\bibnamefont {Kelly}},
  \bibinfo {author} {\bibfnamefont {A.}~\bibnamefont {Megrant}}, \bibinfo
  {author} {\bibfnamefont {A.}~\bibnamefont {Veitia}}, \bibinfo {author}
  {\bibfnamefont {D.}~\bibnamefont {Sank}}, \bibinfo {author} {\bibfnamefont
  {E.}~\bibnamefont {Jeffrey}}, \bibinfo {author} {\bibfnamefont {T.~C.}\
  \bibnamefont {White}}, \bibinfo {author} {\bibfnamefont {J.}~\bibnamefont
  {Mutus}}, \bibinfo {author} {\bibfnamefont {A.~G.}\ \bibnamefont {Fowler}},
  \bibinfo {author} {\bibfnamefont {B.}~\bibnamefont {Campbell}}, \bibinfo
  {author} {\bibfnamefont {Y.}~\bibnamefont {Chen}}, \bibinfo {author}
  {\bibfnamefont {Z.}~\bibnamefont {Chen}}, \bibinfo {author} {\bibfnamefont
  {B.}~\bibnamefont {Chiaro}}, \bibinfo {author} {\bibfnamefont
  {A.}~\bibnamefont {Dunsworth}}, \bibinfo {author} {\bibfnamefont
  {C.}~\bibnamefont {Neill}}, \bibinfo {author} {\bibfnamefont
  {P.}~\bibnamefont {O'Malley}}, \bibinfo {author} {\bibfnamefont
  {P.}~\bibnamefont {Roushan}}, \bibinfo {author} {\bibfnamefont
  {A.}~\bibnamefont {Vainsencher}}, \bibinfo {author} {\bibfnamefont
  {J.}~\bibnamefont {Wenner}}, \bibinfo {author} {\bibfnamefont {A.~N.}\
  \bibnamefont {Korotkov}}, \bibinfo {author} {\bibfnamefont {A.~N.}\
  \bibnamefont {Cleland}},\ and\ \bibinfo {author} {\bibfnamefont {J.~M.}\
  \bibnamefont {Martinis}},\ }\bibfield  {title} {\bibinfo {title}
  {Superconducting quantum circuits at the surface code threshold for fault
  tolerance},\ }\href {https://doi.org/10.1038/nature13171} {\bibfield
  {journal} {\bibinfo  {journal} {Nature}\ }\textbf {\bibinfo {volume} {508}},\
  \bibinfo {pages} {500} (\bibinfo {year} {2014})}\BibitemShut {NoStop}%
\bibitem [{\citenamefont {Motzoi}\ \emph {et~al.}(2011)\citenamefont {Motzoi},
  \citenamefont {Gambetta}, \citenamefont {Merkel},\ and\ \citenamefont
  {Wilhelm}}]{motzoi2011optimal}%
  \BibitemOpen
  \bibfield  {author} {\bibinfo {author} {\bibfnamefont {F.}~\bibnamefont
  {Motzoi}}, \bibinfo {author} {\bibfnamefont {J.~M.}\ \bibnamefont
  {Gambetta}}, \bibinfo {author} {\bibfnamefont {S.~T.}\ \bibnamefont
  {Merkel}},\ and\ \bibinfo {author} {\bibfnamefont {F.~K.}\ \bibnamefont
  {Wilhelm}},\ }\bibfield  {title} {\bibinfo {title} {Optimal control methods
  for rapidly time-varying hamiltonians},\ }\href
  {https://doi.org/10.1103/PhysRevA.84.022307} {\bibfield  {journal} {\bibinfo
  {journal} {Phys. Rev. A}\ }\textbf {\bibinfo {volume} {84}},\ \bibinfo
  {pages} {022307} (\bibinfo {year} {2011})}\BibitemShut {NoStop}%
\bibitem [{\citenamefont {van Dijk}\ \emph {et~al.}(2019)\citenamefont {van
  Dijk}, \citenamefont {Kawakami}, \citenamefont {Schouten}, \citenamefont
  {Veldhorst}, \citenamefont {Vandersypen}, \citenamefont {Babaie},
  \citenamefont {Charbon},\ and\ \citenamefont
  {Sebastiano}}]{PhysRevApplied.12.044054}%
  \BibitemOpen
  \bibfield  {author} {\bibinfo {author} {\bibfnamefont {J.}~\bibnamefont {van
  Dijk}}, \bibinfo {author} {\bibfnamefont {E.}~\bibnamefont {Kawakami}},
  \bibinfo {author} {\bibfnamefont {R.}~\bibnamefont {Schouten}}, \bibinfo
  {author} {\bibfnamefont {M.}~\bibnamefont {Veldhorst}}, \bibinfo {author}
  {\bibfnamefont {L.}~\bibnamefont {Vandersypen}}, \bibinfo {author}
  {\bibfnamefont {M.}~\bibnamefont {Babaie}}, \bibinfo {author} {\bibfnamefont
  {E.}~\bibnamefont {Charbon}},\ and\ \bibinfo {author} {\bibfnamefont
  {F.}~\bibnamefont {Sebastiano}},\ }\bibfield  {title} {\bibinfo {title}
  {Impact of classical control electronics on qubit fidelity},\ }\href
  {https://doi.org/10.1103/PhysRevApplied.12.044054} {\bibfield  {journal}
  {\bibinfo  {journal} {Phys. Rev. Appl.}\ }\textbf {\bibinfo {volume} {12}},\
  \bibinfo {pages} {044054} (\bibinfo {year} {2019})}\BibitemShut {NoStop}%
\bibitem [{\citenamefont {Landi}\ \emph {et~al.}(2012)\citenamefont {Landi}
  \emph {et~al.}}]{landi2012characterization}%
  \BibitemOpen
  \bibfield  {author} {\bibinfo {author} {\bibfnamefont {C.}~\bibnamefont
  {Landi}} \emph {et~al.},\ }\bibfield  {title} {\bibinfo {title}
  {Characterization of arbitrary waveform generator by low resolution and
  oversampling signal acquisition},\ }\href
  {https://doi.org/10.1016/j.measurement.2012.07.016} {\bibfield  {journal}
  {\bibinfo  {journal} {Measurement}\ }\textbf {\bibinfo {volume} {45}},\
  \bibinfo {pages} {2498} (\bibinfo {year} {2012})}\BibitemShut {NoStop}%
\bibitem [{\citenamefont {Singh}\ \emph {et~al.}(2023)\citenamefont {Singh},
  \citenamefont {Zeier}, \citenamefont {Calarco},\ and\ \citenamefont
  {Motzoi}}]{singh2023compensating}%
  \BibitemOpen
  \bibfield  {author} {\bibinfo {author} {\bibfnamefont {J.}~\bibnamefont
  {Singh}}, \bibinfo {author} {\bibfnamefont {R.}~\bibnamefont {Zeier}},
  \bibinfo {author} {\bibfnamefont {T.}~\bibnamefont {Calarco}},\ and\ \bibinfo
  {author} {\bibfnamefont {F.}~\bibnamefont {Motzoi}},\ }\bibfield  {title}
  {\bibinfo {title} {Compensating for nonlinear distortions in controlled
  quantum systems},\ }\href {https://doi.org/10.1103/PhysRevApplied.19.064067}
  {\bibfield  {journal} {\bibinfo  {journal} {Phys. Rev. Appl.}\ }\textbf
  {\bibinfo {volume} {19}},\ \bibinfo {pages} {064067} (\bibinfo {year}
  {2023})}\BibitemShut {NoStop}%
\end{thebibliography}
\end{document}